\documentclass[zconf]{zeus_paper}
\usepackage[english]{babel}
\begin{document}
\newcommand{\pom}{{\rm I\! P}}
\newcommand{\reg}{{\rm I\! R}}
\newcommand{\lsim}{\raisebox{-.6ex}{${\textstyle\stackrel{<}{\sim}}$}}
\newcommand{\gsim}{\raisebox{-.6ex}{${\textstyle\stackrel{>}{\sim}}$}}
\def\coll{Collaboration}
\def\etal{et al.}
\pagenumbering{arabic} 
\pagestyle{plain}
\begin{flushright}
BRIS/HEP/2000-05
\end{flushright}
\begin{center}
{\LARGE Low-x Physics}                                                       
                    
Brian Foster,\\H.H. Wills Physics Lab, University of
Bristol, U.K. \& DESY, Hamburg, Germany \\
%
{\it Invited talk given at `The Quark Structure of Matter'\\
The Royal Society, London, May 24 -- 25, 2000}
\end{center}
\vspace{1cm}
\begin{center}
Abstract
\end{center}
\vspace{0.5cm}
{\small{\it Low-x physics is reviewed, with particular emphasis on
searches for deviations from GLAP evolution of the parton
densities. Although there are several intriguing indications,
both in HERA and Tevatron data, as yet there is no unambiguous
evidence for other than standard next-to-leading-order GLAP
evolution. The framework of dipole models and saturation of
parton densities is examined and confronted with the data. 
Although such models give a good qualitative description of the
data, so do other, more conventional, explanations.}}

\section{Introduction}
\label{sec-int}

The scattering of energetic `simple' particles from an 
unknown target to elucidate its structure is an experimental technique with a long and distinguished history. 
Such scattering experiments have revolutionised our view of the microscopic world; the prototype, and 
most famous, is the scattering of alpha particles from a thin gold foil carried 
out by a previous president of this society, Lord Rutherford, working with 
Geiger and Marsden in
Manchester in 1909. This experiment led to the concept of the nuclear 
atom~\cite{pm:21:669,prslon:82:495}. A similar revolution occurred after the SLAC 
experiments, as 
discussed by Taylor at this meeting~\cite{misc:taylor:rs2000}. These led to the 
general acceptance of the concept of quarks as actual constituents of the 
nucleon, rather than as mathematical abstractions.

\begin{figure}[h]
\begin{center}
\epsfig{file=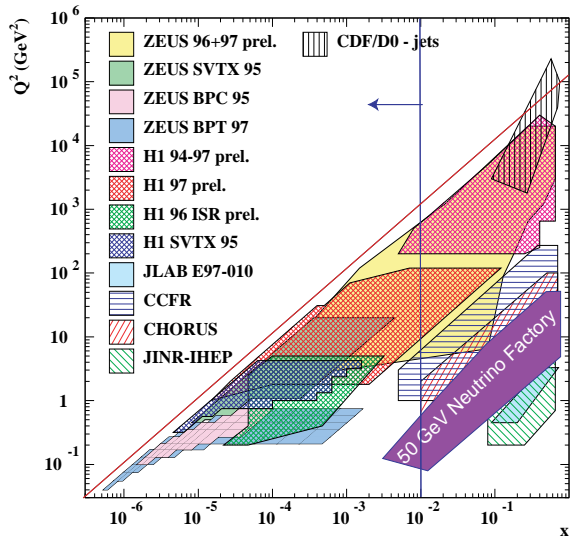,%
      width=8cm,%
      height=8cm%
        }
\end{center}
\caption{The kinematic plane in $x$ and $Q^2$ for all experiments, probing
the parton distribution of the proton. The region of interest in this talk
is indicated to the left of the vertical line. The other line at
close to 45$^o$ indicates the approximate kinematic limit at HERA}
\label{fig:q2xkinreg}
\end{figure}

With the advent of the HERA electron-proton collider, the explorable phase space
in the kinematic invariants $Q^2$ (the virtuality of the exchanged virtual 
photon)
and $x$ (the fractional momentum of the parton involved in the scattering) has 
increased by approximately three orders of magnitude in each variable (see
figure~\ref{fig:q2xkinreg}). 
This 
increase in kinematic range has opened up a new branch of studies,
which can be described generically as `low-$x$ physics'. As will be
seen in this talk, the study of this kinematic region, which for convenience
will be defined by $x < 10^{-2}$, is full of interest and has already led
to many advances in the understanding of the theory of the strong interaction,
Quantum Chromodynamics (QCD). 

Although the study of diffractive processes is intimately linked to many aspects
of low-$x$ physics, constaints of time mean that it is not covered in this talk.
The contribution by J. Dainton~\cite{misc:dainton:rs2000} to this meeting touches 
upon diffraction to some degree. 

\subsection{Theoretical background}
\label{sec:theory:intro}
The interest in low-$x$ physics is that particles with small $x$ are the result of a 
large number of QCD branching processes. 
The behaviour of partons at low $x$ thus reflects the 
dynamics of QCD and allows the behaviour of its couplings and interactions to be 
probed over a large range in the kinematic variables. In particular, 
the evolution of the number of partons as a function of $x$ and $Q^2$ will
be sensitive, depending on the kinematic range,  to the various  
approximations that describe QCD evolution.

Several of the contributions to this meeting have discussed the subject of QCD 
evolution in some depth~\cite{misc:altarelli:rs2000,misc:dokshitzer:rs2000}, 
and there are of course many excellent overviews of the 
subject~\cite{ellis:1996:qcd}, so that is appropriate here only to give a very 
brief summary of the most important points relevant at low $x$.

One of the most important properties of QCD, without which its usefulness
as a theory would be extremely limited, is that of factorisation. This states 
that hard processes can be regarded as a convolution of a `sub-process'
cross section that can be calculated in terms of point-like interactions
together with the probability to find the participating particles in the
target and in the probe. The subsequent hadronisation of the participants in
the hard collision, together with the target and probe remnants, 
can be regarded as an approximately independent process. Thus 
the cross section can be written schematically as:
\begin{equation}
\sigma \sim f_t \otimes f_p \otimes \hat{\sigma}
\label{eq:factorisation}
\end{equation}
where $\hat{\sigma}$ is the sub-process cross section and $f_t, f_p$ are
the parton distribution functions for the target and probe, respectively.
One of the most important results of the factorisation hypothesis is that
the parton distribution functions (PDFs) measured in one process can be used
in the cross-section determination for a completely different
process. Furthermore, QCD provides the tools by which to extrapolate from the 
PDFs measured at one scale to very different scales. 

Specialising now to deep inelastic scattering (DIS), the
PDF for the highly virtual photon can normally
be considered to be a $\delta$ function, so that 
equation~\ref{eq:factorisation} becomes:  
\begin{equation}
\sigma \sim f \otimes \hat{\sigma}
\label{eq:DIS:factorisation}
\end{equation}
where $f$ now represents the PDF of the proton. It
is conventional to assume that
$f$  
satisfies the schematic equation:
\begin{equation}
\frac{\partial f}{\partial \ln \mu^2} \sim
\frac{\alpha_s(\mu^2)}{2\pi}
\cdot \left(f \otimes {\cal P} \right)
\label{eq:RGE}
\end{equation}
where $\mu$ represents the renormalisation scale and ${\cal P}$ is a 
`splitting 
function' that describes the probability of a given parton splitting
into two others. This equation is known
as the (Dokshitzer)-Gribov-Lipatov-Altarelli-Parisi 
equation~\cite{sovjnp:20:95,sovjnp:15:438,np:b126:298,jetp:46:641}.
There are four distinct Altarelli-Parisis (AP) splitting functions 
representing the 4 possible $1 \rightarrow 2$ splittings and referred to as 
$P_{qq}, P_{gq}, P_{qg}$ and $P_{gg}$.  
The calculation of the splitting functions in perturbative QCD in
equation~\ref{eq:RGE} requires approximations, both in 
order of terms which can be taken into account as well as the most important
kinematic variables. The generic form for the splitting functions can be
shown to be~\cite{ellis:1996:qcd}:
\begin{equation}
x{\cal P}(x,\alpha_s) = \sum^\infty_{n=0} \left(\frac{\alpha_s}{2\pi}\right)^n
\left[\sum^n_{m=0} A^{(n)}_m \left\{\ln \left(\frac{1}{x}\right)\right\}^m
+ x\overline{\cal P}^{(n)}(x)\right]
\label{eq:general-splitting-fn}
\end{equation}
where $\alpha_s$ is the strong coupling constant, $\overline{\cal P}^{(n)}(x)$
are the $x$-finite parts of the
AP splitting functions and $A^{(n)}_m$ are 
numerical coefficients that can be calculated, at least in principle, 
for each splitting function. 
The AP splitting functions sum
over terms proportional to $(\alpha_s \ln Q^2)^n$ 
in the perturbative expansion.
Thus, for example, the term in
equation~\ref{eq:general-splitting-fn} 
with $n=m=0$ 
when added to $\overline{\cal P}^{(0)}(x)$ corresponds to leading
order in so-called Gribov-Lipatov-Altarelli-Parisi 
(GLAP) evolution. 
In some kinematic regions, and in particular at low $x$, it must become
essential to sum leading terms in $\ln 1/x$ independent of the value of
$\ln Q^2$. These terms in some sense correspond to 
corrections taking into account so-called 
Balitsky-Fadin-Kuraev-Lipatov~\cite{pl:b60:50,jetp:44:433,jetp:45:199,sovjnp:28:822} 
(BFKL) evolution,
which governs the evolution in $x$ at fixed $Q^2$.
As $x$ falls, this must at some point drive parton evolution. One of
the continuing themes of low-$x$ physics, as will be discussed in 
sections~\ref{sec:otherprobes} and \ref{sec:interpretation}, is the search for 
experimental effects that can be unambiguously attributed to BFKL evolution.

\begin{figure}[h]
\begin{center}
\epsfig{file=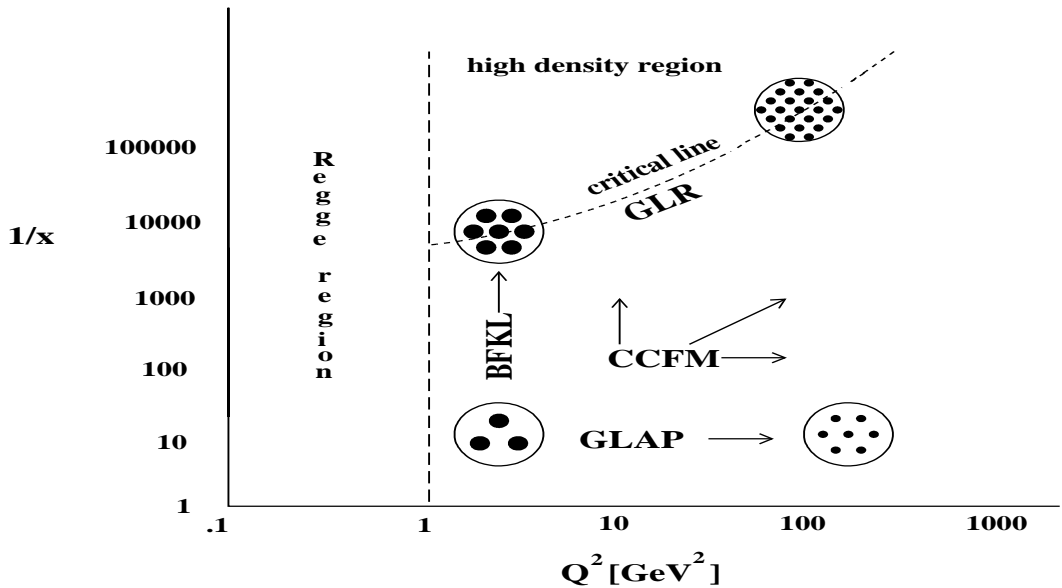,%
      width=14cm,%
      height=8cm%
        }
\end{center}
\caption{Schematic diagram showing different regions of the $\ln 1/x$ and
$\ln Q^2$ plane and the evolution equations expected to hold therein.
The line marked 'saturation' represents the boundary between GLAP
evolution and evolution governed by the GLR equation. The `size' of
partons is also indicated in differing kinematic regions.}
\label{fig:evolution}
\end{figure}

Figure~\ref{fig:evolution} shows the $\ln 1/x$ - $\ln Q^2$ plane at HERA, together
with schematic indications of the directions in which GLAP and BFKL dominate the 
evolution of parton distributions.
The third direction on the figure, labelled `CCFM', refers to an approach
to an integrated evolution containing both the leading 
GLAP and BFKL terms to equal
order developed by
Ciafaloni, Catani, Fiorani and 
Marchesini~\cite{np:b296:49,pl:b234:339,np:b336:18}. 
Also indicated on the 
figure are schematic
indications of both the `size'  
and density of partons in the proton in different kinematic regions.
The transverse size of the partons which can be resolved by a probe
with virtuality $Q^2$ is proportional to $1/Q$, so that the area of the
partonic `dots' in figure~\ref{fig:evolution} falls as $Q^2$ rises.
For particular combinations of parton size and density, the proton will
eventually become `black' to probes, or, equivalently, the component gluons
will become so dense that they will begin to recombine.
The dotted line labelled `Critical line - GLR' refers to the boundary
beyond which it is expected that such parton saturation effects will become
important, i.e. the region in which partons become so densely crowded
that interactions between them reduce the growth in parton density predicted by 
the linear GLAP and BFKL evolution equations. The parton evolution in this 
region can be described by the 
Gribov-Levin-Riskin~\cite{prep:100:1,np:b268:427} 
equation, which explicitly takes into account an absorptive term in the gluon 
evolution equation. Naively, it can be assumed~\cite{quadt:phd:1997} that 
the gluons inside the proton each occupies on average a 
transverse area of
$\pi Q^{-2}$ so that the total transverse area occupied by gluons is
proportional to the number density multiplied by this area, i.e.\
$\pi Q^{-2}xg(x,Q^2)$. Since, as will be discussed later, the gluon density
increases quickly as $x$ falls, and the gluon `size' increases as $Q^{-1}$, 
in the region in which both $x$ and $Q^2$ are small, saturation
effects ought to become important. This should occur when the size
occupied by the partons becomes similar to the size of the proton:
\begin{equation}
xg(x,Q^2)\frac{\pi}{Q^2} = \pi R^2
\label{eq:satlimit}
\end{equation}
where $R$ is the radius of the proton, ($\sim 1$~fm $\sim 5$~GeV$^{-1}$). 
The measured values of
$xg(x,Q^2)$ imply that saturation ought to be observable at
HERA~\cite{proc:ringberg:1999:levin} at low $x$ and $Q^2$, 
although the values of $Q^2$ which satisfy equation~\ref{eq:satlimit}
are sufficiently small that possible non-perturbative and 
higher-twist effects certainly complicate the situation. Of course,
it is also possible that the assumption of homogenous gluon density
is incorrect; for example, the gluon density may be larger in the
close vicinity of the valence quarks, giving rise to
so-called `hot spots'~\cite{pr:189:267}, which could lead to 
saturation being observable at smaller distances and thereby larger $Q^2$. 
The concepts of `shadowing' or saturation have been discussed now
for many
years~\cite{np:b268:427,arevns:44:199,np:b437:107,np:b461:512,np:b493:305,np:b510:355,pr:d49:2233,pr:d50:2225,pr:d53:458,pr:d59:014014,pr:d59:034007,pr:d55:5414,pr:d52:3809,pr:d52:6231,pr:d54:5463,pr:d55:5445,np:b529:451,np:b507:367,pl:b379:239,np:b493:354,pl:b425:369,np:b539:535,zfp:c49:607,pl:b326:161,zfp:c53:331,pr:d60:074012,hep-ph-0007257,np:b537:477,np:b558:285,hep-ph-9911289,proc:ringberg:1999:levin}. 
As will be seen
in section~\ref{sec:sat}, HERA does indeed provide 
data of relevance to such discussions.         
\section{The Structure Function Data}
\label{sec:sfd}
In this section, the most recent structure function data from ZEUS and H1
are presented and discussed. After some initial definitions of kinematic 
variables and the structure functions relevant at low $x$, the data
on $F_2$ are shown and indirect methods of extracting the longitudinal structure 
function, $F_L$ are discussed.
\subsection{Kinematics and structure function formulae}
\label{sec:kin}
The scattering of a lepton from a proton at sufficiently large $Q^2$ can be 
viewed as the elastic scattering of the lepton from a quark or antiquark inside 
the proton. As such the process can be fully described by two relativistic 
invariants.  If the initial (final) four-momentum of 
the lepton is $k (k')$, the initial four-momentum of the  proton is $P$, the 
fraction of the proton's  
momentum carried by the struck quark is $x$ and the final four-momentum  
of the hadronic system is $P'$, the following invariants may be  
constructed: 
\begin{eqnarray} 
s &=& (P+k)^2\\ 
Q^2 &=& -q^2 = -(k' - k)^2 \\ 
y &=& \frac{P \cdot q}{P \cdot k} \\ 
W^2 &=& (P')^2 = (P+q)^2 
\end{eqnarray} 
Energy-momentum conservation implies that:  
\begin{equation} 
x = \frac{Q^2}{2 P \cdot q} 
\end{equation}
so that, ignoring the masses of the lepton and proton:  
\begin{eqnarray}
y &=& \frac{Q^2}{sx}
\label{eq:yq2sx} \\
W^2 &=& Q^2\frac{1-x}{x} \sim \frac{Q^2}{x}  
\label{eq:w2q2x}
\end{eqnarray} 
where the approximate relationship will in general be sufficiently accurate
for the values of $x$ of interest in this talk.  
Since DIS at a given $s$ can be specified by any two of these 
invariants, the most convenient may be chosen, normally $x$ and $Q^2$. 

Equation~\ref{eq:general:sigma} shows the 
general form for the spin-averaged neutral current differential cross section 
in terms of the structure functions ${ F}_1$, ${ F}_2$ and  
${F}_3$:    
\begin{eqnarray}  
\frac{d^2 \sigma}{dx dQ^2}  
& = & \frac{2\pi \alpha^2}{x Q^4}   
\left[2xy^2{ F}_1 + 2(1-y)  { F}_2 \right. \nonumber \\ 
&\; \pm & \left. \{ 1-(1-y)^2 \} x { F}_3 \right]  
\label{eq:general:sigma} 
\end{eqnarray} 
where the $+$ sign in the $\pm$ term 
is taken for $e^-$ and the $-$ sign for $e^+$
interactions. The structure functions are products of 
quark distribution functions 
and the couplings of the current mediating the interaction. They are in  
general functions of the two invariants required to describe the interaction.  
 
To leading order in the QCD-improved parton model,
in which quarks are massless, 
have spin $\frac{1}{2}$ and  
in which they develop no $p_T$,  the Callan-Gross  
relation~\cite{prl:22:156}: 
\begin{eqnarray} 
2 x F_1 (x) & = & F_2 (x)
\label{eq:CallanGross} 
\end{eqnarray} 
is satisfied. At the next order, $p_T$ must be taken into  
account and this relation is violated. This is usually quantified by  
defining a  
longitudinal structure function, $F_L$, such that   
$F_L = F_2 - 2x F_1$. 
Substituting into equation~\ref{eq:general:sigma} gives: 
\begin{eqnarray}  
\frac{d^2 \sigma} 
{dx dQ^2} & = & \frac{2\pi \alpha^2}{x Q^4}   
 \left[   
Y_+ \cdot
 F_2(x, Q^2) \right. \nonumber \\
 & \; - & \left.  {y^2} F_L(x, Q^2) + Y_- \cdot x F_3(x, Q^2) 
\right] 
\label{eq:Fl:sigma} 
\end{eqnarray}
where $Y_{\pm}$ are kinematic factors given by: 
\begin{equation}
Y_{\pm} = 1 \pm (1-y)^2
\label{eq:Y}
\end{equation}
At low $x$, in general $Q^2 \ll M_Z^2$, $xF_3$ vanishes and 
equation~\ref{eq:Fl:sigma} reduces to that for
photon exchange. In the rest of this talk, electroweak effects will be 
neglected.

In general, the form of the structure functions beyond 
leading order depends on 
the renormalisation and factorisation scheme used.  
In the so-called `DIS' scheme, 
the  logarithmic singularity  
produced by collinear gluon emission is absorbed into 
the definition of the quark distribution, so that 
the structure functions have a particularly simple form and can be expressed to
all orders as 
\begin{eqnarray} 
 F_2(x, Q^2) & = & \sum_{i=u,d,s,c,b} A_i(Q^2) \left[ xq_i(x,Q^2) + 
   x\overline{q}_i(x,Q^2) \right]  
\label{eq:F2:qpm}  
\end{eqnarray} 
The parton distributions $q_i(x,Q^2)$ and $\overline{q}_i(x,Q^2)$ refer  
to quarks and antiquarks of type 
$i$. The quantities
$A_i(Q^2)$ are given by the square of the electric charge of quark or 
antiquark 
$i$. However, in the $\overline{MS}$ scheme, 
the form of $F_2$ changes with order in QCD. In LO QCD it has the form:
\begin{eqnarray} 
 F_2(x, Q^2) & = & \sum_{i=u,d,s,c,b} A_i(Q^2)  
\int_{x}^{1} 
   \frac{dy}{y} \left(\frac{x}{y}\right) \left[ 
\left\{ \delta(1 - \frac{x}{y})  
+ \frac{\alpha_s}{2 \pi} C^{\overline{\mbox{\tiny {MS}}}}_q 
\left(\frac{x}{y}\right)\right\} \right. \nonumber \\
&\cdot & \left.     
\left( yq_i(y,Q^2) + 
   y\overline{q}_i(y,Q^2) \right)  +  \left\{ \frac{\alpha_s}{2 \pi} 
C^{\overline{\mbox{\tiny{MS}}}}_g 
\left(\frac{x}{y}\right) \right\} yg(y,Q^2) \right] 
\label{eq:F2:MSbar} 
\end{eqnarray} 
where  $g(x,Q^2)$ is the gluon density in the proton, 
$\alpha_s(Q^2)$ is the QCD running coupling constant
and $C_q(x)$ and $C_g(x)$ are scheme-dependent `coefficient 
functions'. 

In contrast, the longitudinal structure function contains 
no collinear divergence at first-order in QCD so that:  
\begin{eqnarray}
 F_L (x, Q^2) & = &    
 \frac{ \alpha_{s}(Q^2)} {2\pi} \sum_{i=u,d,s,c,b} A_i(Q^2) 
\left\{ \frac{4}{3}  
\int_{x}^{1}  
   \frac{dy}{y} \left( \frac{x}{y} \right) ^2 \left[ yq_f(y,Q^2) + 
   y\overline{q}_f(y,Q^2) \right] \right.
   \nonumber \\ 
 & + & \left.  2 \int_{x}^{1} \frac{dy}{y} \left(\frac{x}{y} \right)^2  
\left( 1 - \frac{x}{y} \right)yg(y,Q^2) \right\}  
\label{eq:Fl} 
\end{eqnarray}
independent of the factorisation scheme employed.  

Taking account of radiative corrections via the term $\delta_r$, 
equation~\ref{eq:Fl:sigma} becomes:
\begin{eqnarray}  
\frac{d^2 \sigma} 
{dx dQ^2} & =& \frac{2\pi \alpha^2}{x Q^4} \cdot (1 + \delta_r)  \cdot 
\left[  Y_+ \cdot 
 F_2(x, Q^2) 
  - {y^2} F_L(x, Q^2) 
 \right] 
\label{eq:F2} 
\end{eqnarray}
A useful quantity known as the `reduced cross section' can be defined from 
equation~\ref{eq:F2} by taking the kinematic factors to the left-hand side,i.e.:
\begin{eqnarray}  
\frac{x Q^4}{2\pi \alpha^2 Y_+ \cdot (1 + \delta) }\frac{d^2 \sigma} 
{dx dQ^2} & =& \sigma_r \nonumber \\
&=& 
 F_2(x, Q^2) 
  - \frac{y^2}{Y_+} F_L(x, Q^2) 
\label{eq:redsigma} 
\end{eqnarray}
and, provided $y$ is small, $\sigma_r$ is to a good 
approximation equal to $F_2$.

An alternative formalism to describe DIS interactions at low $x$ in terms of 
total virtual photon-proton cross sections is particularly useful when 
discussing the low-$Q^2$ region and the transition to real photoproduction.
The total cross section can be written as the sum of the cross sections
for transversely and longitudinally polarised virtual photons:
\begin{equation}
\sigma^{\gamma^*p}_{\rm tot}(W^2,Q^2) \equiv \sigma_T + \sigma_L 
\label{eq:sigmagp}
\end{equation}
where $x$ and $W$ are related using equation~\ref{eq:w2q2x}. This
leads directly to expressions for $F_2$ and $F_L$ in terms of virtual photon-proton cross sections:
\begin{eqnarray}
\sigma^{\gamma^*p}_{\rm tot}(W^2,Q^2) \sim 
&F_2& = \frac{Q^2}{4\pi^2\alpha}(\sigma_T + \sigma_L)
\label{eq:f2siglt}\\
&F_L& = \frac{Q^2}{4\pi^2\alpha}\sigma_L
\label{eq:flsigl}
\end{eqnarray}

\subsection{The $F_2$ data at `medium' $Q^2$}
\label{sec:f2med}
In this section the most recent $F_2$ data at `medium' values of
$Q^2$ from the two HERA experiments are discussed. The term `medium $Q^2$'
used here is essentially a definition related to the characteristics
of the H1 and ZEUS detectors; the term covers the structure function
measurements in which the minimum scattered electron angle (and hence
the minimum $Q^2$)
that can be measured is determined by the dimensions of the main
calorimeters of the two experiments. This is in contrast to the
`low-$Q^2$' region, where, as discussed in section~\ref{sec:f2low},
the minimum $Q^2$ range is determined by the geometric acceptance of 
small, special-purpose detectors placed downstream
in the electron-beam direction.

The measurement of $F_2$ is a very complex and painstaking effort and has
been often described before~\cite{ijmp:a13:1543,ijmp:a13:3385}. Since 
both the H1 and ZEUS detectors are sufficiently 
hermetic, the two invariants required to define the event kinematics fully
can be reconstructed from measurements on the 
electron, on the hadronic final state corresponding to the struck quark, or 
on mixtures of the two. The optimal method depends on the kinematic  
region of interest and on the properties and resolutions of the detectors. The 
size of the radiative corrections is also very dependent on the reconstruction 
method employed. The basic measurements made are the angles and energies  
of the electron and current jet. Many methods have
been used by the two experiments, of which probably the most important 
for the measurement of $F_2$ are: the Electron Method, in which the energy
and the angle of the scattered electron are used; the 
Double Angle method~\cite{proc:heraworkshop:1991:23},
in which the angles of the scattered electron and the current jet are used;
the $\Sigma$ method~\cite{nim:a361:197}, which uses the current jet and electron 
energies and the electron angle, and the $P_T$ method~\cite{zfp:c72:399}, which 
uses the Double Angle method with the additional constraint of $p_T$ balance. 

In order to determine $F_2$ 
the following steps are carried out.  
First, a sample of DIS events is selected,  
basically by requiring an identified electron in the detector.  
The kinematic variables are reconstructed using one
of the methods discussed above.
The data are binned in $x$ and $Q^2$  
with bin sizes determined by detector resolution,  
statistics, migration in and out of the 
bin due to the finite resolution of the experiments, etc.   
Estimates of the background in each bin are made and 
statistically subtracted. The background in the low-$x$ region is dominated
by photoproduction processes in which a fake electron is reconstructed
because of confusion with hadronic debris from photoproduction interactions,
which of course have a much higher cross section than DIS processes.    
The data are corrected for acceptance, radiative effects and 
migration via Monte Carlo simulation.   
Multiplying the corrected number of events by the 
appropriate kinematic factors shown in  
equation~\ref{eq:F2} and subtracting an estimate for $F_L$ 
gives values of $F_2$ and hence the quark and antiquark 
densities in the chosen bins. 
There are great gains in physics terms to be made by pushing the precision of 
these measurements to the limits. Since for much of the available phase space 
systematic effects are dominant, this requires progressively better
understanding of the detectors involved and simulation of their response to
parts per mille.    
    
\begin{figure}[h]
\begin{center}
\epsfig{file=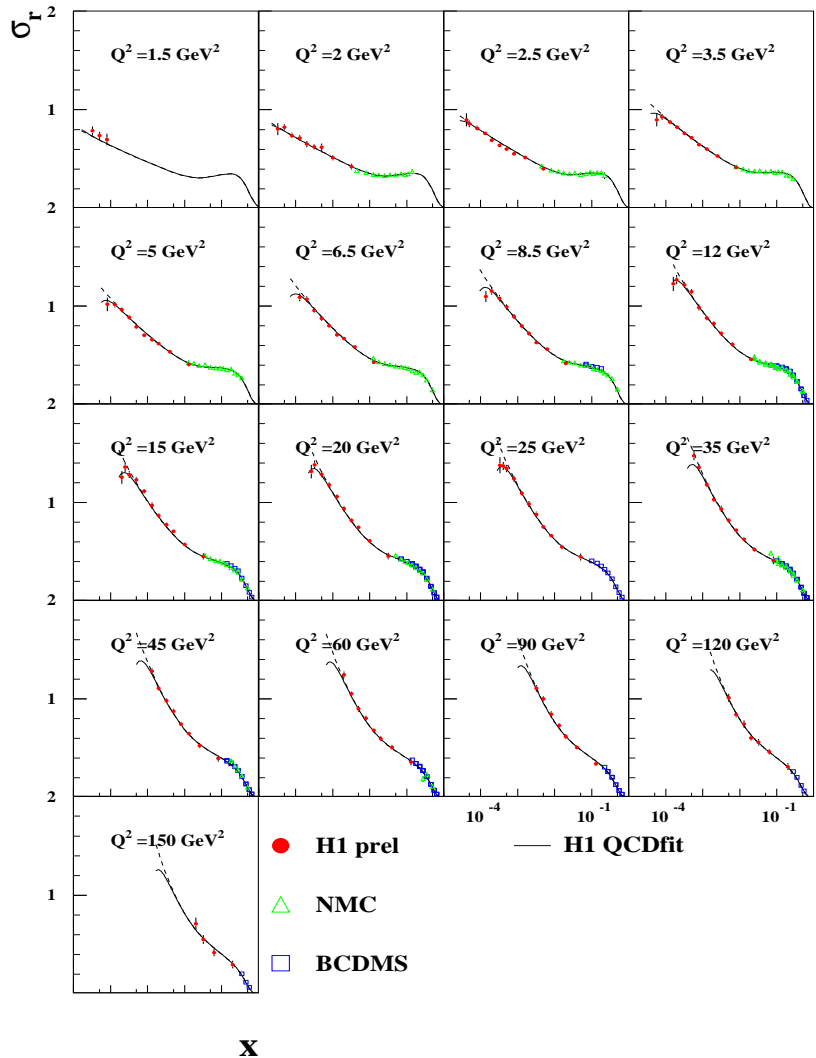,%
      width=11cm,%
      height=15cm%
        }
\end{center}
\caption{The preliminary H1 data on the reduced cross section
(proportional to $F_2$) from
the 1996-97 data-taking period. Also shown are points from the fixed-target
experiments NMC (triangles) and BDCMS (squares). The solid curve shows
the NLO QCD fit carried out by H1, while the dotted curve
visible at the lowest $x$ corresponds to the expectation
for $F_L = 0$, as discussed in the text.}
\label{fig:F2H1:Q2bins}
\end{figure}

Figure~\ref{fig:F2H1:Q2bins} 
shows the preliminary H1 measurement of the 
reduced cross section (see equation~\ref{eq:redsigma})
in bins of $Q^2$ as a function of $x$. Also shown are data
from the fixed-target experiments NMC~\cite{np:b483:3} and
BCDMS~\cite{pl:b223:485,pl:b237:592}. The bins from $Q^2$ = 1.5 GeV$^2$
to 150 GeV$^2$ are shown. In the relatively small
region of overlap, there is good agreement between the H1 and fixed-target
data. The most obvious characteristic of the data is the
steep rise of $F_2$ at low $x$. The curve shown on the figure is the result
of an NLO QCD fit by the H1 collaboration to this data together
with earlier measurements at higher $x$ from 
NMC. The NLO QCD fit~\cite{misc:kleinlps99}, based on
GLAP evolution, uses three light flavours with charm added via the boson-gluon 
fusion process and uses $\alpha_s(M^2_Z) = 0.118$. It gives an excellent 
fit to the data over the full kinematic range. The quality of the QCD
fit permits the conclusion that the rise of $F_2$ at low $x$ is unambiguously
associated with a dramatic rise in the
gluon distribution. Also shown in 
figure~\ref{fig:F2H1:Q2bins} as the dashed line is the expectation of
the fit for $F_2$ alone. As can be seen, there is a small departure from
the measured value of $\sigma_r$, implying that in this kinematic region, the
effect of $F_L$ begins to become perceptible; this is discussed further in
section~\ref{sec:Fl}.

\begin{figure}[h]
\begin{center}
\epsfig{file=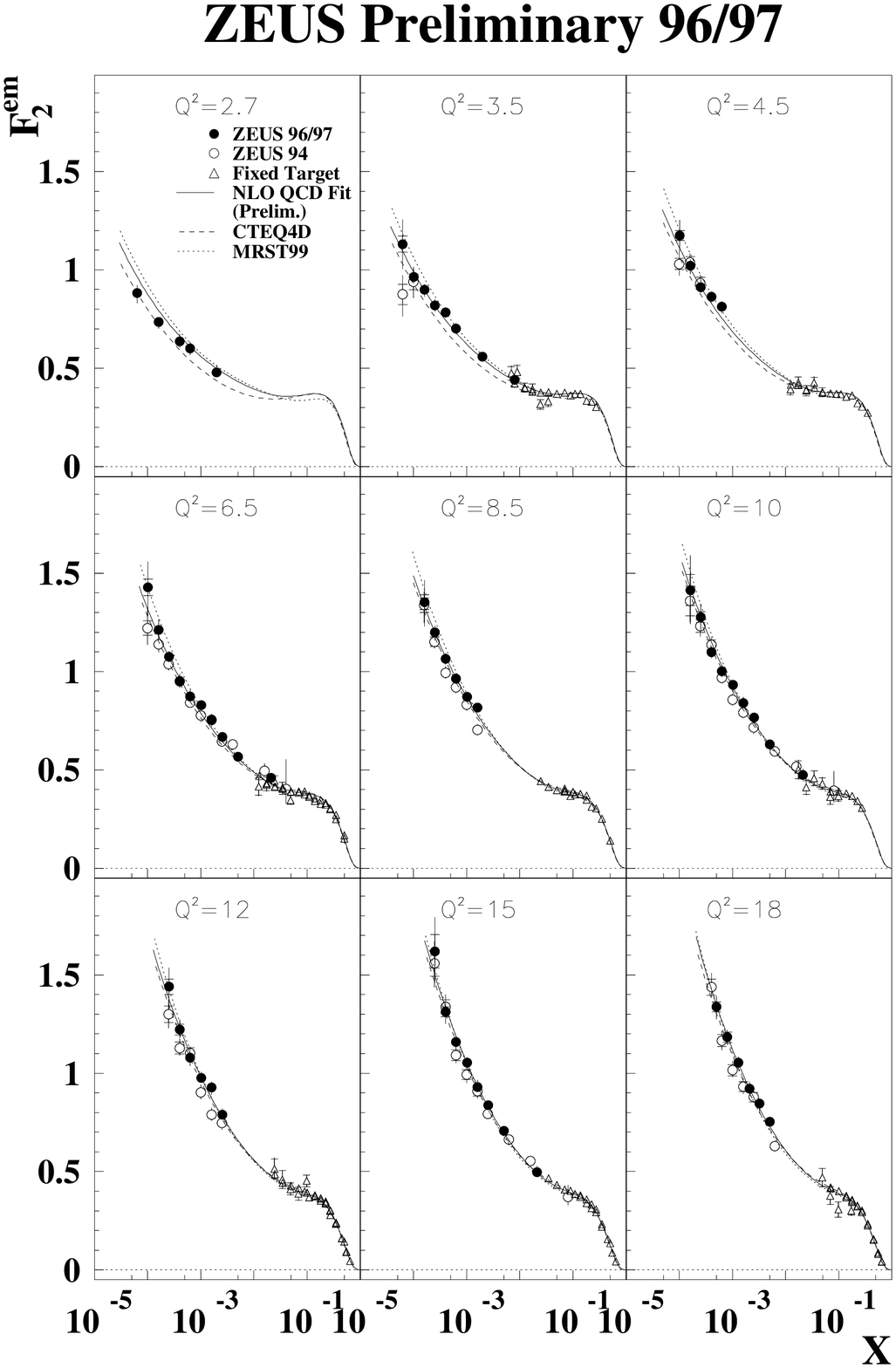,%
      width=10cm,%
      height=9cm%
        }
\end{center}
\caption{The lowest $Q^2$ bins of the preliminary ZEUS data on $F_2$ from
the 1996-97 data-taking period. The preliminary data
are in good agreement with the published data (shown as open circles).
Also shown are points from fixed-target
experiments. The solid curve shows
the ZEUS NLO QCD fit, while the dashed line shows the CTEQ4D curve
and the dotted line that from MRST99.}
\label{fig:F2ZEUS:Q2bins}
\end{figure}

Figure~\ref{fig:F2ZEUS:Q2bins} 
shows the preliminary 1996-97 ZEUS data, plotted as the $F_2$ structure function as a function of $x$ in $Q^2$ bins with the QCD $F_L$ prediction subtracted. Only the lowest $Q^2$ bins are shown; $F_2$ has been determined up to
$Q^2$ = 30,000 GeV$^2$. Fixed target data from 
BCDMS~\cite{pl:b223:485,pl:b237:592}, E665~\cite{pr:d54:3006},
NMC~\cite{np:b483:3}, and SLAC~\cite{pl:b282:475} are also shown.    
The data agree well with the H1 data and show the same dominant feature of a 
very steep rise at low $x$. The data are also very well described by the NLO QCD 
global fit to parton distributions of CTEQ4D~\cite{pr:d55:1280} and MRST99\cite{epj:c14:133}.
  
\begin{figure}[h]
\begin{center}
\epsfig{file=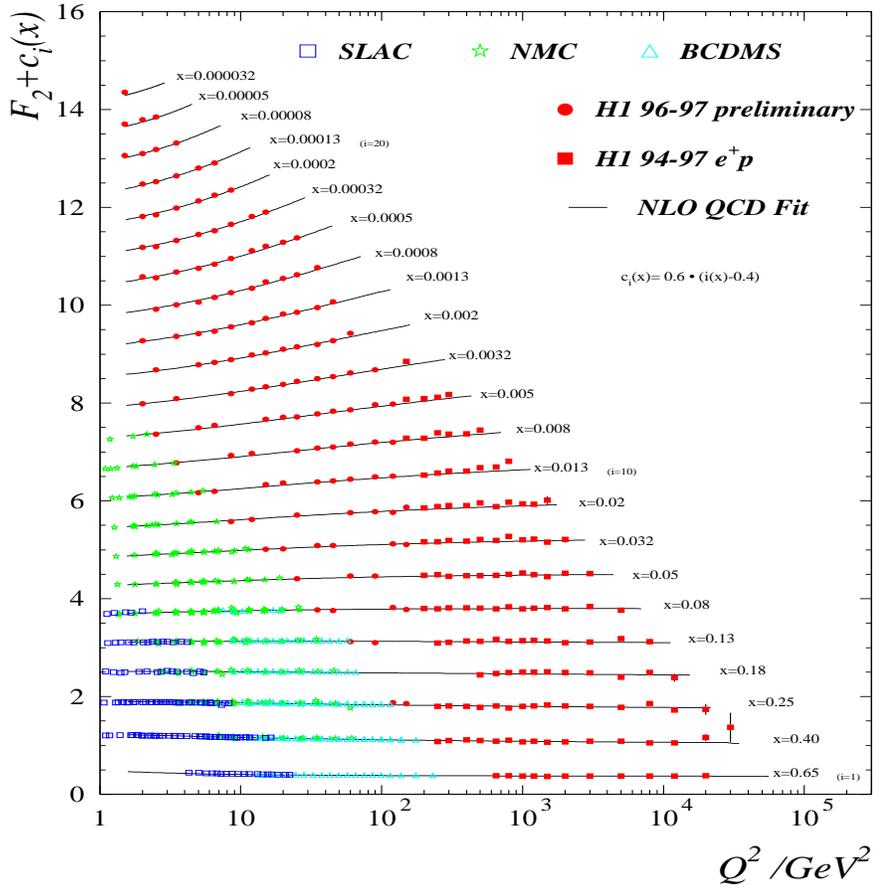,%
      width=12cm,%
      height=12cm%
        }
\end{center}
\caption{Preliminary H1 data on $F_2$ from 1996-97 and published
data from 1994-97 in bins of $x$ as a function of $Q^2$. Also
plotted are fixed-target data from SLAC, NMC and BCDMS. Each
$x$ bin is offset by the amount indicated in the legend for
ease of visibility. The curves show the H1 NLO QCD fit to the data.
The flat 'scaling' regime at high $x$ gives way at lower $x$ to
steep scale breaking due to gluon radiation.}
\label{fig:F2H1:xbins}
\end{figure}

Figure~\ref{fig:F2H1:xbins} shows the H1 data together with
data from NMC and BCDMS, now plotted in
$x$ bins as a function of $Q^2$. 
The data cover approximately five orders of magnitude in
both $x$ and $Q^2$. At high $x$, approximate scaling in $Q^2$ can be clearly
observed. As $x$ falls, deviations from scaling become stronger and
stronger. The lines on the figure are the result of the NLO QCD fit,
which can again be seen to give an excellent description of the data. 

\begin{figure}[h]
\begin{center}
\epsfig{file=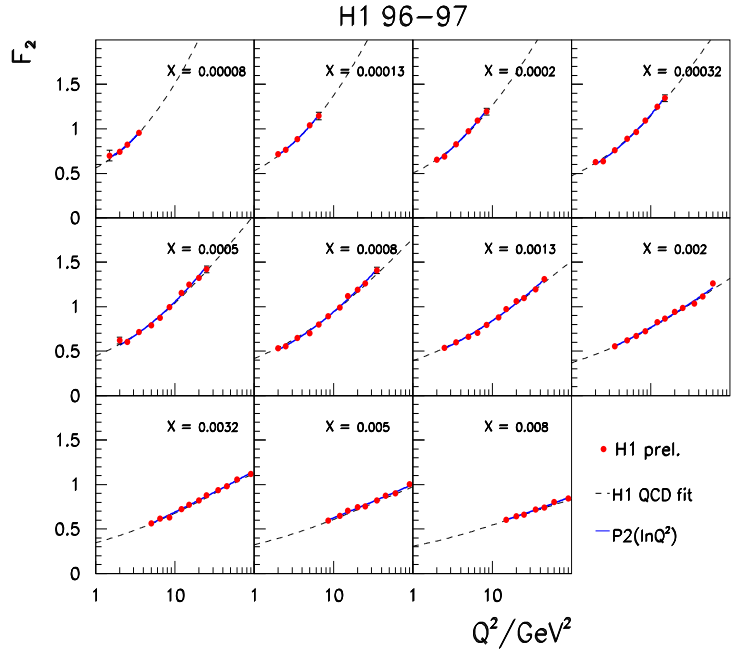,%
      width=10cm,%
      height=9cm%
        }
\end{center}
\caption{Preliminary H1 data on $F_2$ from 1996-97 in bins of $x$
as a function of $Q^2$. The solid line shows a fit to the form of
equation~\protect\ref{eq:F2polyfit}, while the dashed line shows
the H1 NLO QCD fit.}
\label{fig:H1F2polyfit}
\end{figure}

The H1 data for $F_2$ at low $x$ in $x$ bins as a function of 
$\ln Q^2$ is shown in figure~\ref{fig:H1F2polyfit}. 
It can clearly be seen that the data
are not linear in $\ln Q^2$. In fact they fit well to a second-order polynomial
of the form
\begin{equation}
F_2 = A(x) + B(x) \ln Q^2 + C(x) (\ln Q^2)^2
\label{eq:F2polyfit}
\end{equation}
and this polynomial fit is almost indistinguishable from the H1 NLO QCD fit,
except at the highest $Q^2$. 

\begin{figure}[h]
\begin{center}
\epsfig{file=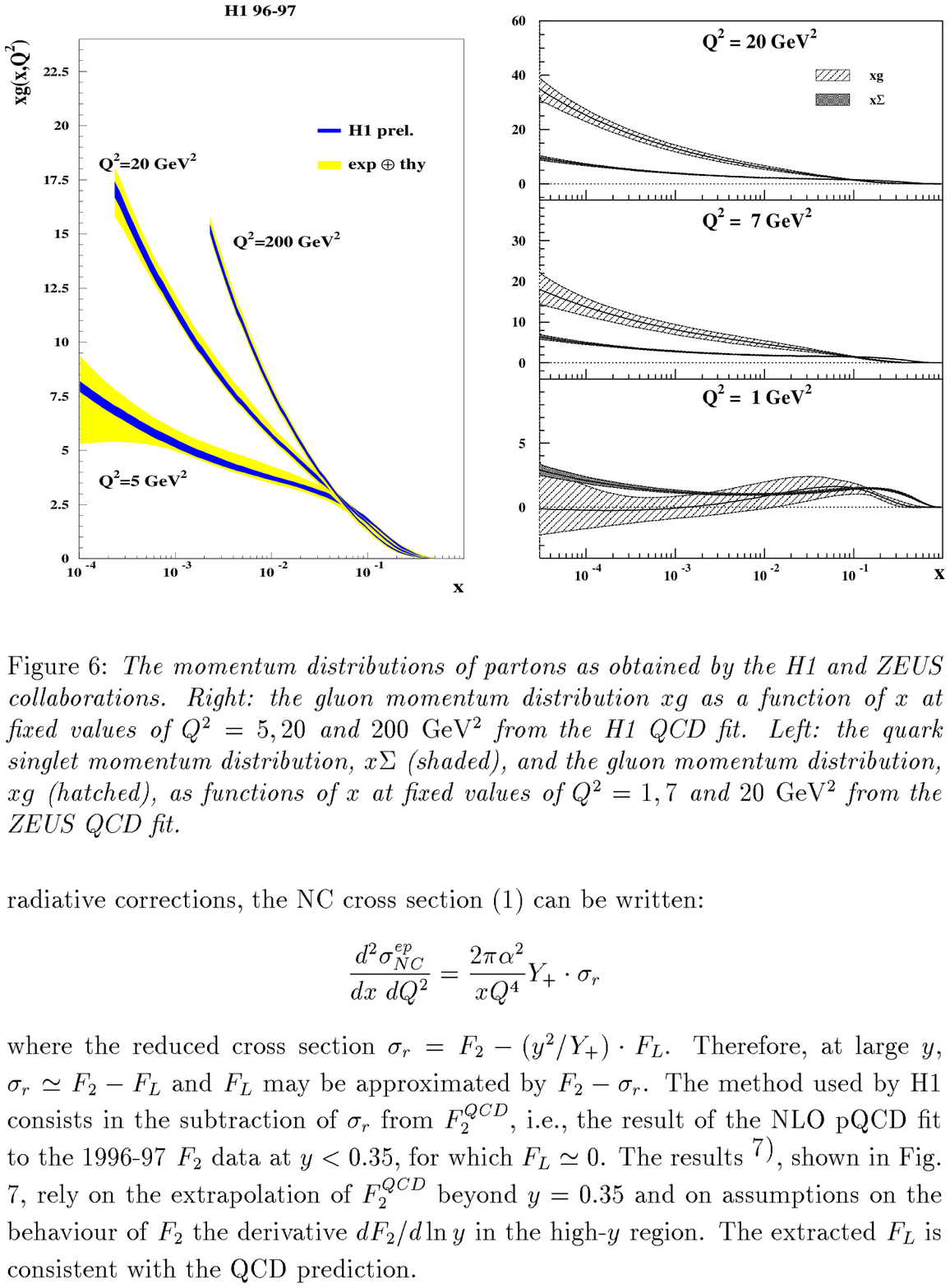,%
      width=12cm,%
      height=6cm,%
      clip=%
        }
\end{center}
\caption{The left-hand plot shows preliminary H1 data on determination
of the gluon from $F_2$ data from 1996-97.
The gluon density is shown for three values of $Q^2$;
the central band shows the statistical uncertainty while the
outer bands show the systematic and the theoretical uncertainties
added in quadrature. The right-hand side plot
shows ZEUS data on determination
of the gluon from $F_2$ data from 1995.
The gluon density is shown for three values of $Q^2$;
all data down to $Q^2 > 1$ GeV$^2$ is included in the 
NLO QCD fit. The bands show all uncertainties
added in quadrature. The light shaded band is the gluon density while
the dark shaded band shows the sea.}
\label{fig:H1ZEUS:gluon}
\end{figure}

The most convenient and useful way to parameterise the deviations of the
data from scaling is to examine the logarithmic derivative, $\partial
F_2/ \partial \ln Q^2$, which in leading-order QCD is directly proportional
to the gluon density at twice the $x$ of the derivative\footnote{It should
be noted that, although a good approximation at LO, this relation
becomes increasingly less valid at higher orders. Although a very
useful qualitative relationship, it should therefore be used with
circumspection.}:
\begin{equation}
\frac{\partial F_2}{\partial \ln Q^2} = \frac{2 \alpha_s}{9 \pi} xg(x, Q^2)
\label{eq:logder}
\end{equation}
Having fit the data to the form of equation~\ref{eq:F2polyfit}, it is
straight-forward to obtain the logarithmic derivative. Since this 
is proportional to the LO
gluon density, it is clear that a precision measurement of the scaling 
violations can be used directly to determine the gluon distribution. 
Figure~\ref{fig:H1ZEUS:gluon} 
shows such determinations from both
H1 and ZEUS. 
The H1 determination comes from the NLO QCD fit referred to
above whereas that from ZEUS comes from an NLO fit to published 
data~\cite{epj:c6:603}. The steep rise in the gluon density as $x$
falls is apparent. Also noticeable, particularly in the ZEUS determination, is 
that this rise becomes weaker and weaker as $Q^2$ falls. Indeed, for 
$Q^2 = 1$ GeV$^2$, the gluon density falls below that of the singlet quark
structure function and is essentially compatible with zero. This
seems to contradict the `standard' picture in which the rise of $F_2$, which
is of course only directly sensitive to the density of charged partons,
is driven by the quark-antiquark pairs produced from gluons. However, these
interesting effects only become obvious at very low values of $Q^2$, and it
is not clear that it makes sense to talk about `gluon densities' at these low
$Q^2$ values\footnote{It was interesting to note
Dokshitzer's comment in the discussion sessions at this meeting that in fact
it does make sense to discuss gluon distributions at such low values of $Q^2$.}.
Nevertheless, it is certainly true that the NLO QCD fits themselves seem to give
a perfectly satisfactory 
description of the general features of the data down to 
these low values of $Q^2$. 
Much more will be said on this subject in section~\ref{sec:sat}.

\begin{figure}[h]
\begin{center}
\epsfig{file=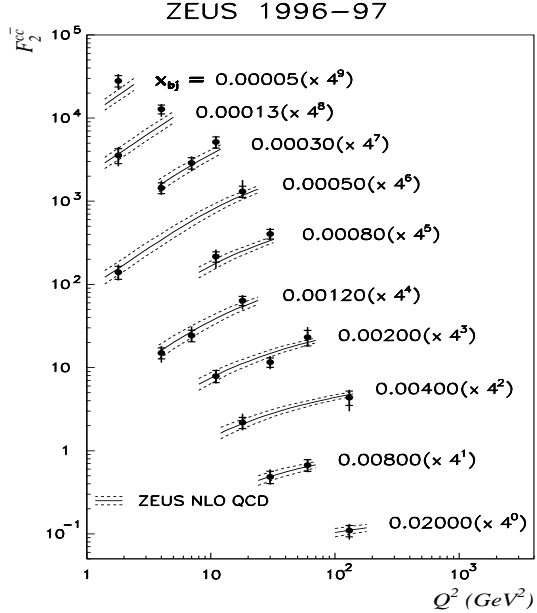,%
      width=7cm,%
      height=8cm%
        }
\end{center}
\caption{ZEUS data on the charm structure
function, in bins of $x$ as a function
of $Q^2$. The band shows the ZEUS NLO QCD fit.} 
\label{fig:ZEUSF2charm}
\end{figure}

\begin{figure}[h]
\begin{center}
\epsfig{file=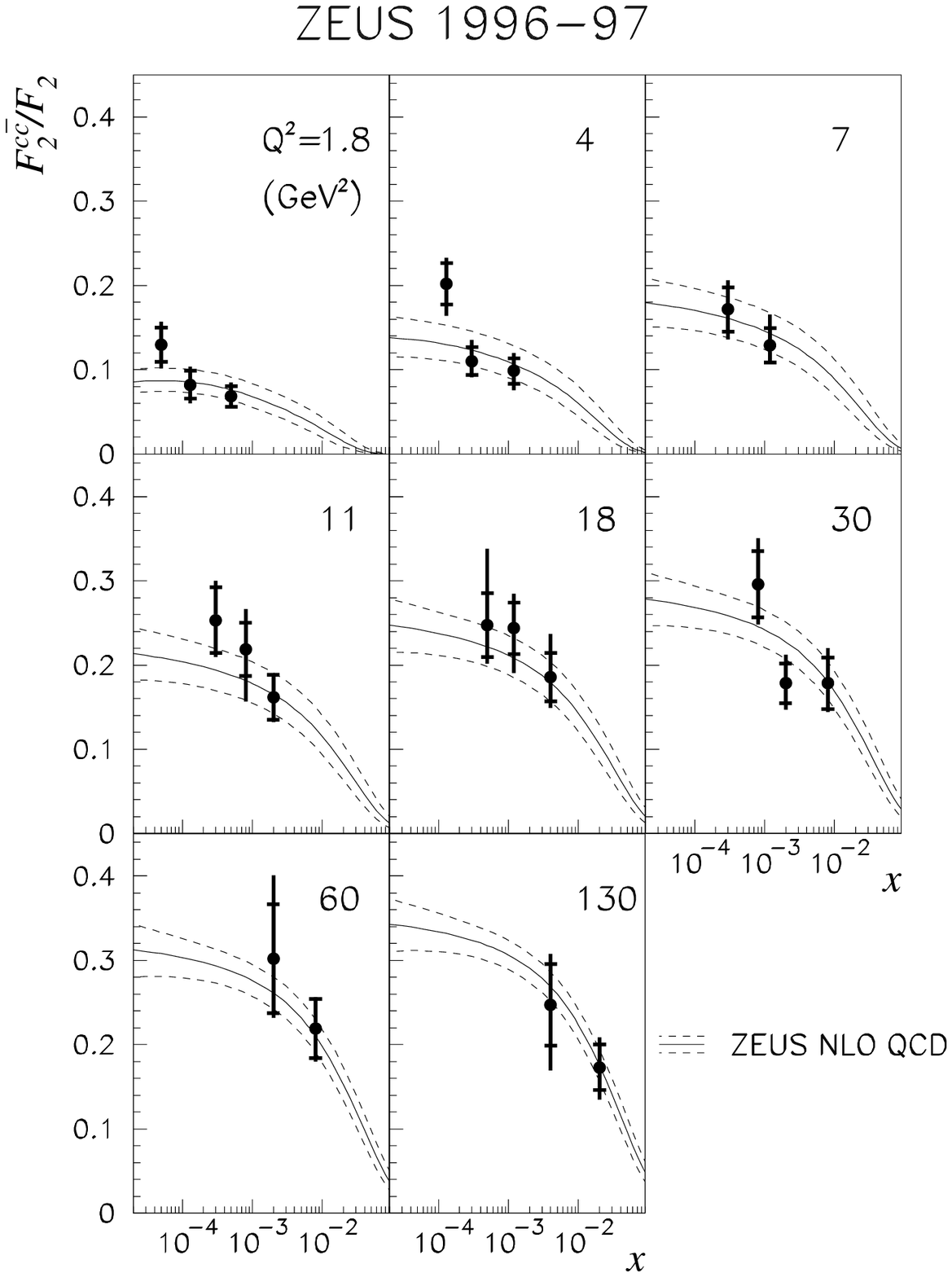,%
      width=9cm,%
      height=8cm%
        }
\end{center}
\caption{ZEUS data on the charm structure
function, plotted as a ratio of the
inclusive structure function, $F_2$ in bins of $Q^2$ as
a function of $x$. The band shows the ZEUS NLO QCD fit.} 
\label{fig:ZEUSF2charm:ratio}
\end{figure}

In addition to the high-precision data on the fully inclusive $F_2$, the ZEUS
collaboration has also presented data on semi-inclusive DIS in which a
charm quark or antiquark is involved in the hard scatter~\cite{epj:c12:35}. 
Figure~\ref{fig:ZEUSF2charm} 
shows the $F_2^{c\overline{c}}$ data in $x$ bins
as a function of $Q^2$. A qualitatively similar pattern of scaling
violations to that
in the fully inclusive $F_2$ can be seen; however, the scaling violations
seem, within the relatively large errors, to be stronger than in
the inclusive case and to set in rather earlier. While part of
this effect can be attributed to the effect of the charm-quark mass, 
it is also to be expected since the
dominant process in DIS charm production is boson-gluon fusion, which is 
entirely driven by the gluon density in the proton. Once again, it can
be seen that the NLO QCD fit gives an excellent description
of the data. Figure~\ref{fig:ZEUSF2charm:ratio}
shows the ratio of the charm over the inclusive $F_2$. At small $x$
the ratio flattens, implying that the charm and inclusive structure
functions grow at the same rate, as is 
to be expected if both are dominated by the
gluon in this region. For low values of $x$, the ratio falls at fixed $x$
as $Q^2$ falls. This is consistent with the observation discussed above that as $Q^2$ falls the gluon density at fixed $x$ also 
falls. 
\subsection{The $F_2$ data at `low' $Q^2$}
\label{sec:f2low}
The ZEUS and H1 detectors are not perfectly hermetic, since it is clearly 
necessary to allow the beams to enter and leave the apparatus. Thus the 
`beam-hole' 
limits the angular acceptance of the detectors both at very forward and 
very backward directions. In the very backward direction (small 
lepton-scattering angles) this limits the $Q^2$ values that can be accurately 
measured to around $\sim 2$ GeV$^2$. In order to access smaller $Q^2$ (and 
thereby smaller $x$), the geometrical acceptance of the detectors must be
extended in some way. There are two main ways in which this has been
achieved. The first is to shift the interaction vertex in the direction of the
proton beam, typically by of order 60 cm, so that the electron has further to
travel before it strikes the rear calorimeters. This means that the
geometrical edge of the detector now corresponds to a smaller scattering angle,
and hence lower $Q^2$ can be accepted. The other method is to install small,
high precision, detectors further upstream in the electron beam direction
which can thereby detect much smaller scattering angles than the main detectors.
Both ZEUS and H1 have such detectors, although so far only ZEUS have published
results. 

\begin{figure}[h]
\begin{center}
\epsfig{file=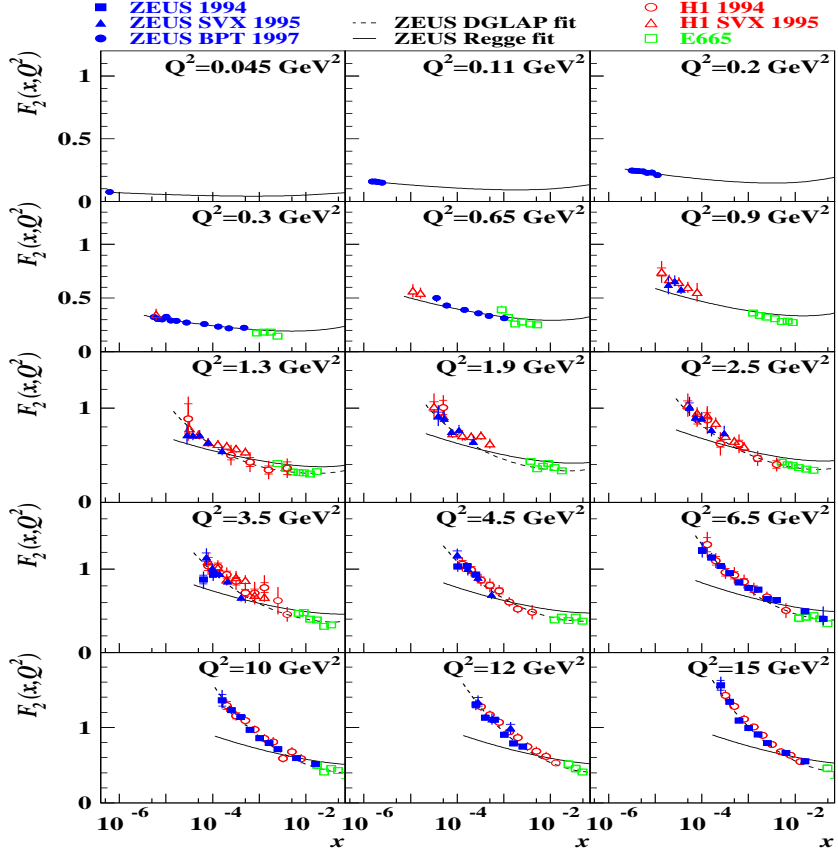,%
      width=11cm,%
      height=12cm%
        }
\end{center}
\caption{ZEUS BPT data on $F_2$ in bins of $Q^2$ as
a function of $x$. Also shown are earlier
ZEUS data as well as data from H1 and E665.
The solid line shows the results of the `ZEUS Regge fit' to
the form of equation~\protect\ref{eq:GVDM+REGGE}, while 
the dotted line shows the result of the ZEUS NLO QCD fit.} 
\label{fig:ZEUSBPT:q2bins}
\end{figure}

ZEUS published some time ago results using their Beam Pipe Calorimeter 
(BPC)~\cite{pl:b407:432}, which is a small tungsten-scintillator sampling 
calorimeter placed 2.94 m
away from the interaction point in the electron beam direction. In 1997, two
silicon-microstrip detector planes were added in front of the calorimeter in 
order to improve the position resolution. ZEUS has recently published the final
results from this BPC/BPT combination~\cite{hep-ex-0005018}, which extend the
measurement of $F_2$ down to $x \sim 6 \cdot 10^{-7}$ and $Q^2 \sim 0.045$ 
GeV$^2$. Figure~\ref{fig:ZEUSBPT:q2bins} shows the final ZEUS data in bins of 
$Q^2$ as a function of $x$. 
The new data match well with the previous ZEUS BPC 
data, as well as with that from other experiments in the overlap region. 
However, the extrapolation of the ZEUS Regge fit (see below) into the 
fixed target regime is generally of order 15\% above this data.
\begin{figure}[h]
\begin{center}
\epsfig{file=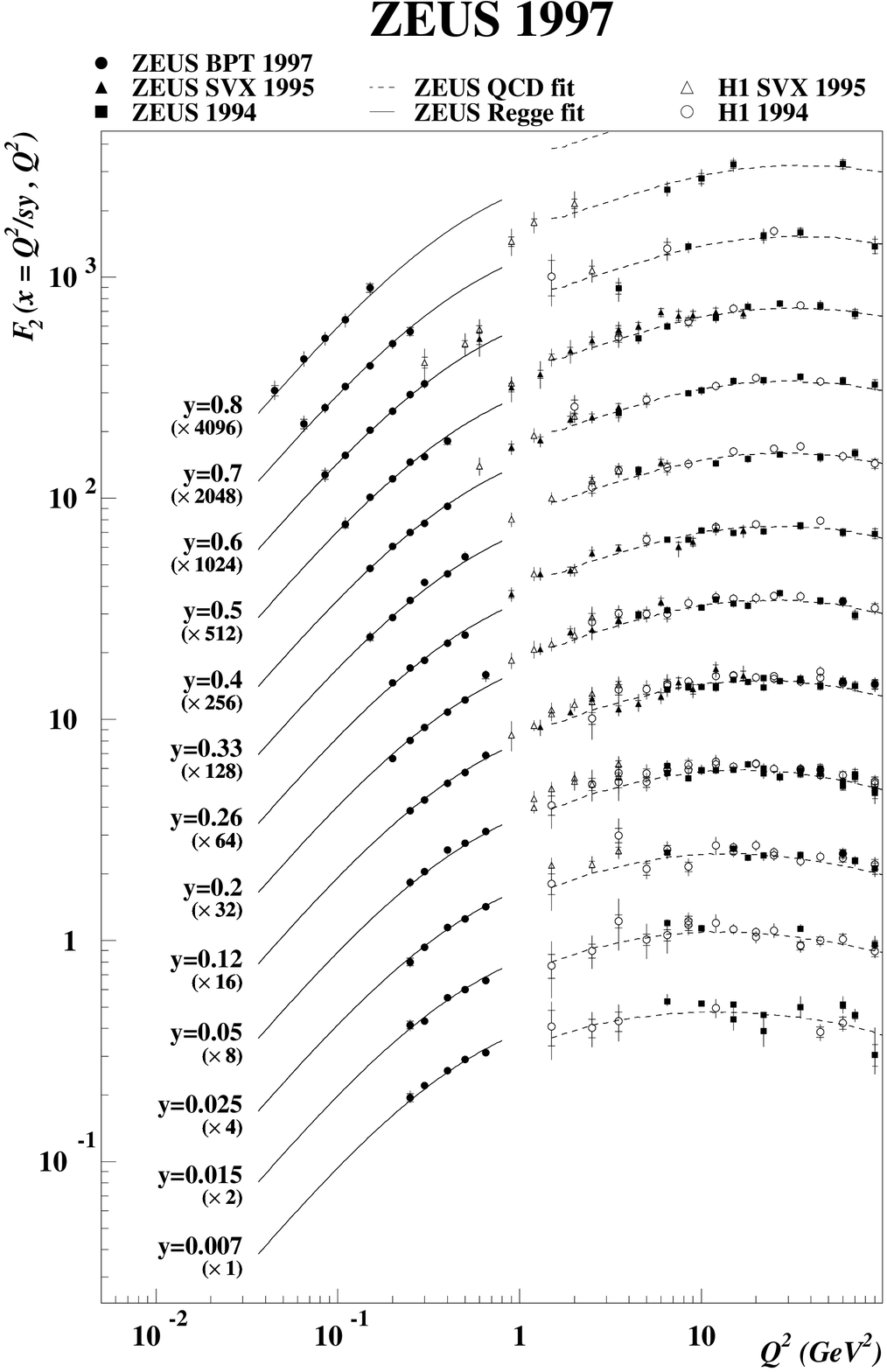,%
      width=9cm,%
      height=10cm%
        }
\end{center}
\caption{ZEUS BPT data on $F_2$ in bins of $y$ as
a function of $Q^2$. Also shown are earlier
ZEUS data as well as data from H1 and E665.
The solid line shows the results of the `ZEUS Regge fit' to
the form of equation~\protect\ref{eq:GVDM+REGGE}, while 
the dotted line shows the result of the ZEUS NLO QCD fit. }
\label{fig:ZEUSBPT:ybins}
\end{figure}

The solid curve labelled `ZEUS Regge fit' on figure~\ref{fig:ZEUSBPT:q2bins}
shows the result of a fit to the form: 
\begin{equation}
F_2(x,Q^2) = \left(\frac{Q^2}{4\pi^2\alpha}\right)\cdot
\left(\frac{M_0^2}{M_0^2+Q^2}\right)\cdot \left(A_\reg\cdot
\left(\frac{Q^2}{x}\right)^{\alpha_\reg-1}+A_\pom\cdot
\left(\frac{Q^2}{x}\right)^{\alpha_\pom-1}\right)
\label{eq:GVDM+REGGE}
\end{equation}
where $A_\reg, A_\pom$ and $M_0$ are constants and $\alpha_\reg $ and
$\alpha_\pom$ are the Reggeon and Pomeron intercepts, respectively.  
This phenomenological
parameterisation is based on the combination of a simplified version
of the generalised vector meson dominance model~\cite{pl:b40:121} for
the description of the $Q^2$ dependence and Regge
theory~\cite{collins:1977:regge} for the description of the $x$ dependence of
$F_2$. Regge Theory is most applicable to the description of cross sections at 
asypmtotic energy. Equations~\ref{eq:w2q2x} and \ref{eq:sigmagp}, which relate
$F_2$ to cross sections evaluated at energies proportional to $x^{-1}$, imply 
that Regge theory should describe the very low-$x$ data well. This is borne
out by figure~\ref{fig:ZEUSBPT:q2bins}, where the ZEUS Regge fit gives a good 
description of the data up to $Q^2 \sim 1$ GeV$^2$. Above this $Q^2$, however, the
Regge description rapidly fails, whereas the ZEUS NLO QCD fit, shown for
$Q^2 > 1$ GeV$^2$, is an excellent description of the data from here to the 
highest $Q^2$. 
  
Figure~\ref{fig:ZEUSBPT:ybins} shows the ZEUS $F_2$ data in bins
of constant $y$ as a function of $\ln Q^2$. 
For $Q^2 \gsim 1$
GeV$^2$, the data are roughly independent of $Q^2$, whereas at lower $Q^2$
they fall rapidly, approaching the $Q^{-2}$ fall-off that would be expected in
the limit $Q^2 \rightarrow 0$ from conservation of the electromagnetic current. 
Whether this dependence indicates that this limit has already been reached, or
whether other effects, for instance saturation, are responsible, will be 
discussed in section~\ref{sec:sat}.

\subsection{The $F_L$ structure function}
\label{sec:Fl}
As discussed in section~\ref{sec:kin}, the differential 
cross section for DIS
at low $x$ depends on two structure functions, $F_2$ and $F_L$. 
Since in principle both $F_2$ and $F_L$
are unknown functions that depend on $x$ and $Q^2$, the only way in which they
can be separately determined is to measure the differential cross section at
fixed $x,Q^2$ and at different values of $y$, since as shown in 
equation~\ref{eq:Fl:sigma}, the effect of $F_L$ is weighted by $y^2$
whereas $F_2$ is weighted by $1 + (1-y)^2$. 
However, since $Q^2 = sxy$, fixed $x$ and $Q^2$ implies 
taking measurements at different values of $s$. This can certainly
in principle be accomplished by reducing the beam energies in HERA.
However, the
practical difficulties for the experiments and the accelerator
inherent in
reducing either the proton or electron beam energy, or both,
by a factor sufficient to permit an accurate measurement of $F_L$ 
mean that it has not to date been attempted. An alternative way to
achieve the same end is to isolate those events in which the incoming
lepton radiates a hard photon in advance of the deep inelastic scattering,
thereby reducing the effective collision energy. Unfortunately, the
acceptance of the luminosity taggers typically used to detect such photons
is sufficiently small and understanding the acceptance sufficiently difficult
that, although both experiments are working on the analysis, neither
has as yet produced results.

In the absence of any direct determination, the H1 experiment has utilised
its ability to detect events at very large values of $y$ in order to
carry out an indirect measurement of $F_L$. The determinations of $F_2$
rely on the fact that most of the measurements are
made at values of $y$ sufficiently small that the
effects of $F_L$ are negligible; those at higher $y$ usually have an
estimate of $F_L$, which according to QCD is in any case normally a small
fraction of $F_2$, subtracted off. The H1 collaboration inverts this
procedure by isolating kinematic regions in which the contribution of $F_L$
is maximised and then subtracts off the QCD prediction of $F_2$ measured
at lower $y$. 

As remarked earlier, figure~\ref{fig:F2H1:Q2bins} shows the reduced 
cross section, defined by equation~\ref{eq:redsigma}, in which
the contribution of $F_L$ can be seen at the lowest $x$ (which,
for fixed $Q^2$, corresponds to the highest $y$) as the difference
between the full QCD fit and that with $F_L$ set to zero. Thus,
$F_L$ can be estimated from the following relationship:
\begin{equation}
F_L = \left( F_2^{\rm QCDfit} -
\frac{xQ^4}{2\pi\alpha^2}\sigma_r \right) 
\cdot \frac{Y_+}{y^2}
\label{eq:flmeas}
\end{equation}
\begin{figure}[h]
\begin{center}
\epsfig{file=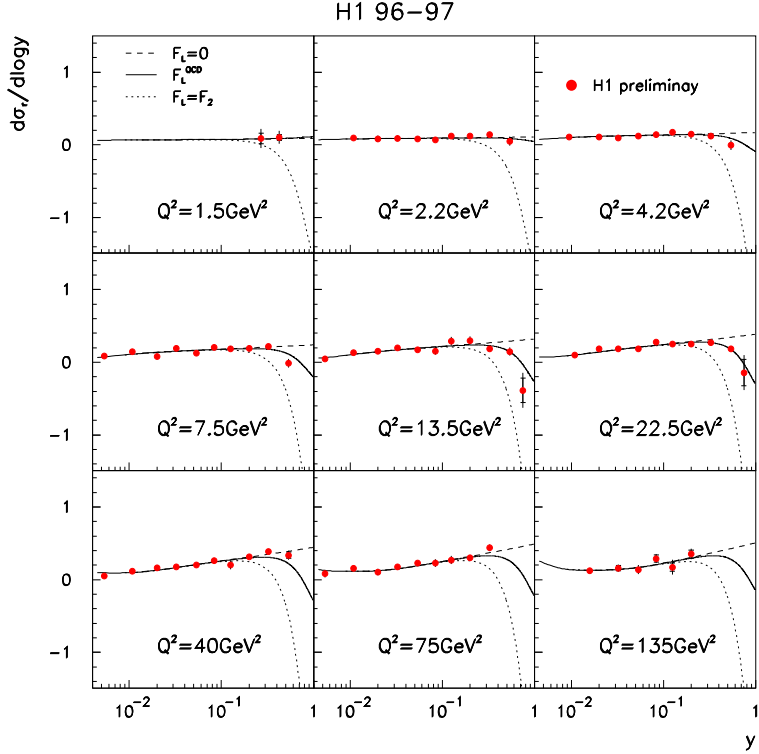,%
      width=10cm,%
      height=10cm%
        }
\end{center}
\caption{Preliminary H1 data on the derivative of the reduced
cross section with respect to $\ln y$ in $Q^2$ bins. The solid
curve shows the result of an NLO QCD fit with the value of
$F_L$ as predicted by QCD from the measured $F_2$. The dotted line
shows the same fit with $F_L=$ and the dashed line shows the fit with
$F_L$
 = $F_2$.}
\label{fig:H1:dsigmardlny}
\end{figure}
\begin{figure}[h]
\begin{center}
\epsfig{file=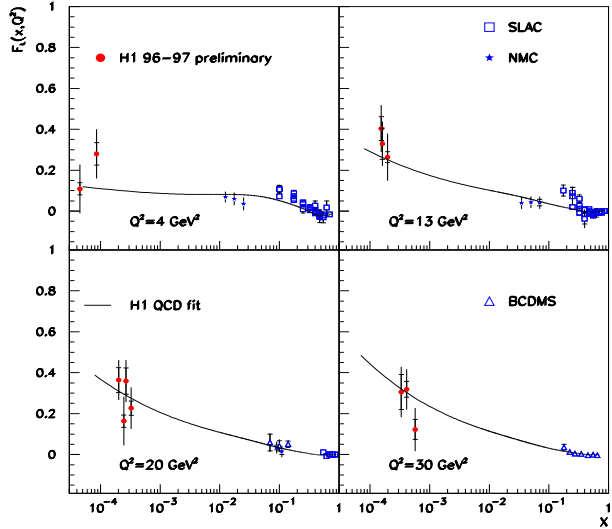,%
      width=8cm,%
      height=7cm%
        }
\end{center}
\caption{Preliminary H1 estimate of $F_L$. The
$F_L$ values obtained are plotted in $Q^2$ bins as a function of $x$.
Also shown are earlier bins at higher $x$ from the SLAC
and NMC experiments. The solid line is the prediction of the H1
NLO QCD fit.}
\label{fig:H1:fl}
\end{figure} 

An alternative method used by H1 employs the derivatives of the
reduced cross section with respect to $\ln y$, thereby making
rather different QCD assumptions. Differentiating
equation~\ref{eq:redsigma} leads to the following expression:
\begin{equation}
\frac{\partial \sigma_r}{\partial \ln y} = \frac{\partial F_2}{\partial \ln y}
- \frac{2y^2(2-y)}{Y^2_+}F_L - \frac{y^2}{Y_+}\cdot 
\frac{\partial F_L}{\partial \ln y}
\label{eq:dsigmardlny}
\end{equation}
which leads to greater sensitivity to $F_L$ via the stronger $y$ dependence
at the cost of involving derivatives of $\sigma_r, F_2$ and $F_L$, the
quantity to be measured. It is instructive to consider various 
restrictions:
\begin{itemize}
\item Small $y$ - here $\partial \sigma_r/\partial \ln y \sim  
\partial F_2/\partial \ln y$. For low $x$, $F_2$ can be well approximated
by: 
\begin{eqnarray}
F_2 &\propto& x^{-\lambda} \propto y^\lambda 
\label{eq:f2xlambda} 
\end{eqnarray}
so that: 
\begin{eqnarray}
\frac{\partial F_2}{\partial \ln y} &=& \lambda y^\lambda 
\label{eq:df2dy}
\end{eqnarray}
which can be expanded as:
\begin{equation}
\frac{\partial F_2}{\partial \ln y} \propto  \lambda e^{\lambda \ln y} \sim
\lambda(1 + \lambda \ln y \ldots )
\label{eq:df2dlny:exp}
\end{equation}
provided $\lambda \ln y$ is small. From this it is clear that $\partial \sigma_r/\partial \ln y$ is linear in
$\ln y$; 
\item $F_L = 0$ - for all $y$, $\partial \sigma_r/\partial \ln y$ is linear in
$\ln y$ for the same reason as above; 
\item $F_L \neq 0$ and large $y$ - $\partial \sigma_r/\partial \ln y$ is 
non-linear in $\ln y$ and the deviations are proportional to $F_L$ and
its logarithmic derivative; 
\item $Q^2$ large at small $y$ - this implies $x$ becoming larger so that
at some point the approximation of equation~\ref{eq:f2xlambda} starts to
fail and therefore there are deviations from non-linearity. 
\end{itemize}
All of these features can be seen in the preliminary H1 data of 
figure~\ref{fig:H1:dsigmardlny}. 
At the largest values of $y$, the
deviation from linearity implies that $F_L$ is non-zero. Although
it is in principle possible to solve the differential equation
for $F_L$ implied by equation~\ref{eq:dsigmardlny}, in practice the
data are insufficiently precise, so that the value of the derivative
is taken from the QCD fit. Variations in this are included in the
systematic error. Also in principle it is possible to
iterate the $F_L$ estimated in this way with that assumed in
the measurement of $F_2$; once again, the precision of the
data does not permit this and in any case the correlation would
be very large.  

The H1 collaboration have used the first method discussed above to estimate
$F_L$ for $Q^2 > 10$ GeV$^2$ and the second for smaller $Q^2$. The results
are shown in figure~\ref{fig:H1:fl}, together with earlier determinations
from SLAC~\cite{pl:b282:475}, NMC~\cite{np:b483:3}
and BCDMS~\cite{pl:b237:592}.
The curve is the result
of an NLO QCD fit to the H1 data deriving from the $F_2$ determination,
i.e. by deriving the 
gluon and quark distributions from scaling violations and then 
calculating $F_L$ using a QCD formula such as equation~\ref{eq:Fl}. The
QCD prediction is in good agreement with the H1 estimate.

In summary, although the indirect determinations of $F_L$ are both
interesting and important, there is no substitute for a direct 
measurement. Since after the HERA upgrade the lowest $Q^2$
regime will no longer be accessible because of the
new final-focus quadrupoles that close off the small scattering angle 
aperture, such data will presumably
have to come from hard initial-state radiation events. First results
from H1 and ZEUS are eagerly awaited.
\section{Other probes of QCD dynamics at small $x$}
\label{sec:otherprobes}
Despite the very high precision and very large kinematic range of the 
structure-function data shown in the previous section, there was no
obvious sign of any deviation from NLO GLAP evolution (although see
section~\ref{sec:sat}). Indeed, 
there are very good reasons why this
should be so, even if BFKL dynamics were 
important~\cite{hep-ph-9912445} in this kinematic range. It
is generally agreed that the chances of observing any deviation from
GLAP evolution are greatly enhanced by examining certain exclusive
processes in particular corners of phase space. Several such processes
are examined in this section: the production of
forward jets and forward $\pi^0$ at HERA and the study of 
dijets with a large separation
in rapidity at the Tevatron.
\subsection{Forward jet production at HERA}
\label{sec:forwardjet}
One of the most marked characteristics of GLAP evolution is that successive
parton branchings are strongly ordered in $k_T^2$, the square of the
transverse momentum of the parton. In the case of BFKL, there is essentially
a random walk in $k_T^2$, while strong ordering occurs in $1/x$. These
observations immediately imply the corner of phase space that is most likely
to exhibit BFKL effects: small x, since this will enhance the importance of
the $\ln (1/x)$ terms in the pQCD expansion, and $p_T$ of the struck
quark $\sim Q^2$, which will strongly suppress GLAP evolution because of the
strong $k_T$ ordering of successive parton branchings between 
the virtual photon
and the struck quark~\cite{npps:18c:125,jp:g17:1443}. The 
kinematic properties of the struck quark can be
reconstructed in several ways, of which the most usual is to tag a high-energy
jet. The kinematic requirements discussed above imply selecting events with
low $x$ and therefore low $Q^2$, so that the balancing jet with comparable $p_T$
will be in the very forward direction.
\begin{figure}[h]
\begin{center}
\epsfig{file=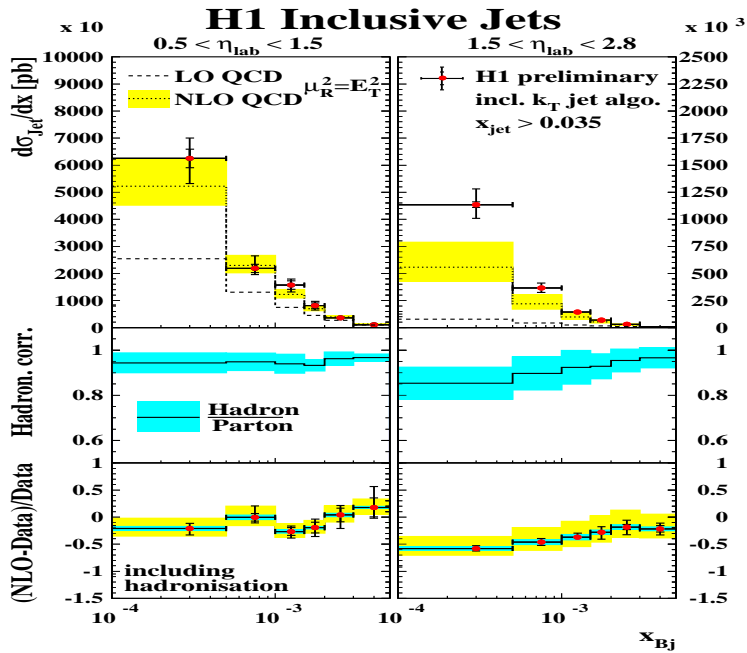,%
      width=10cm,%
      height=10cm%
        }
\end{center}
\caption{Preliminary H1 data on production of inclusive jets, found
with the inclusive $k_T$ jet algorithm, in two bins
of $\eta$ as a function of $x$. The dotted line shows the result of an LO QCD fit, whereas
the dotted line within the shaded band shows the NLO QCD fit at a scale
of $\mu_R^2 = E_T^2$ with an estimate of the theoretical uncertainty.
The central plots show the effect of hadronisation corrections, while
the difference of the data from the NLO predictions is shown at the bottom.}
\label{fig:H1:forwardjetsEt}
\end{figure}

\begin{figure}[h]
\begin{center}
\epsfig{file=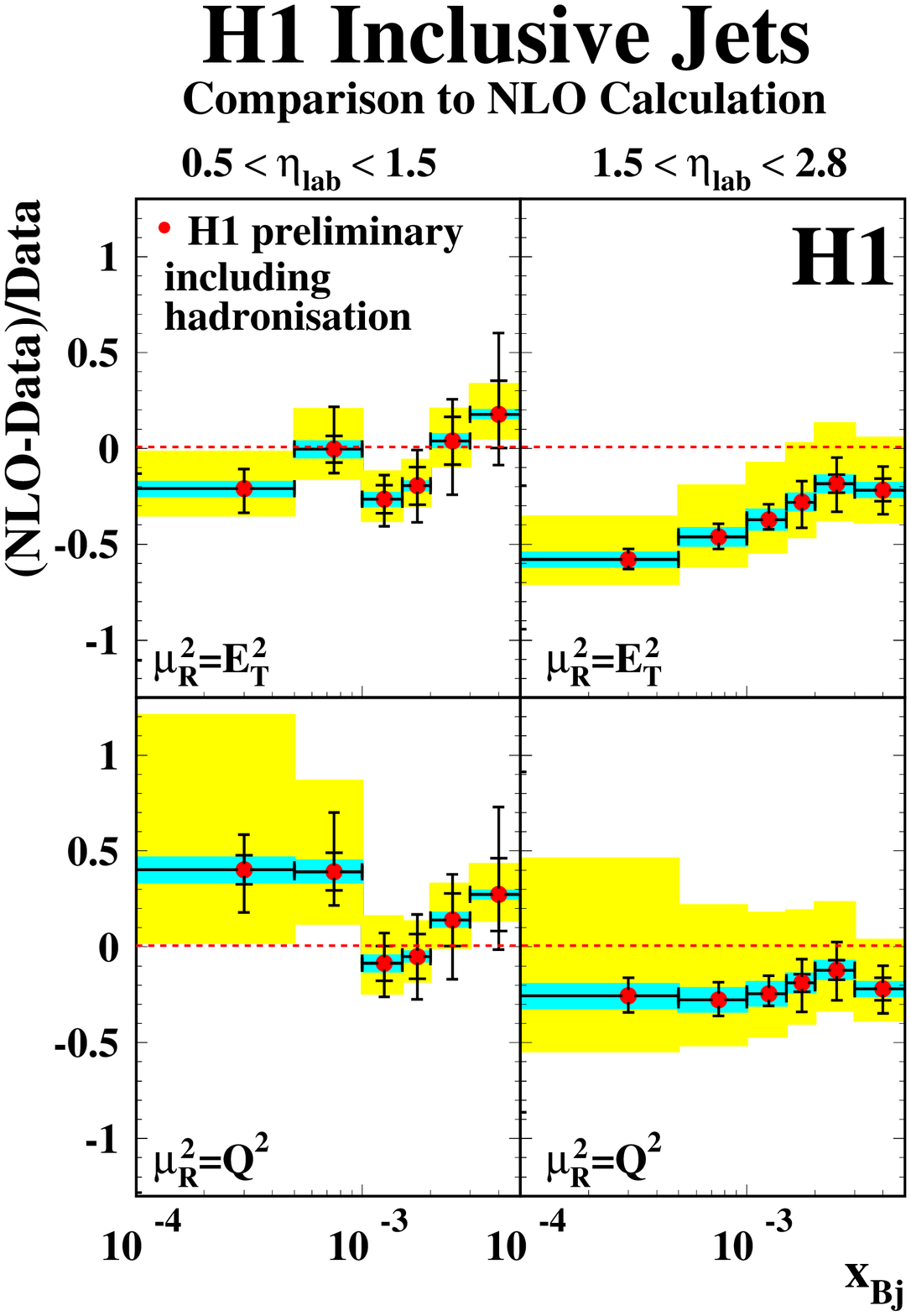,%
      width=10cm,%
      height=10cm%
        }
\end{center}
\caption{Preliminary H1 data on production of inclusive jets, found
with the inclusive $k_T$ jet algorithm, in two bins
of $\eta$ as a function of $x$. The plots show the difference between
the data and the NLO QCD prediction. Two
predictions are compared, one with a normalisation scale $\mu_R^2 = E_T^2$,
the other with $\mu_R^2 = Q^2$. The central bands show the effect of hadronisation corrections, while the remaining theoretical uncertainty is shown as the outer band.}
\label{fig:H1:forwardjetsEtQ2}
\end{figure}

The H1 collaboration presented preliminary results
on forward jet production at the DIS2000 
conference in Liverpool~\cite{misc:schoerner:dis2000}, 
in which they isolated jets
using the inclusive $k_T$ jet algorithm in two pseudorapidity regions:
`central', defined as $0.5 < \eta < 1.5$, and `forward', defined as
$1.5 < \eta < 2.8$. Jets were selected with fractional energy $x_{\rm jet} 
= \frac{E_{\rm jet}}{E_{\rm pbeam}} > 0.035$. The differential cross-section
$d\sigma_{\rm jet}/dx$ for the `forward' and `central' regions is shown
in figure~\ref{fig:H1:forwardjetsEt}. 
Whereas NLO QCD gives a reasonable
description of the data in the central region, it clearly falls below the
data in the forward region at the lowest values of $x$. Although the 
hadronisation corrections are largest in the forward direction, they are insufficient to 
explain the discrepancy. The variation in the NLO QCD prediction by varying the
scale by a factor of two in each direction is also large, but again insufficient
to explain the shortfall. However, $E_T^2$ is used as the hard scale in these
calculations, and it is entirely unclear whether this, or $Q^2$, is the 
appropriate scale. Figure~\ref{fig:H1:forwardjetsEtQ2}
shows a comparison of the
calculations assuming $E_T^2$ and $Q^2$ to give the hard scale in the NLO 
calculation. It can be seen that not only is the discrepancy reduced when $Q^2$ 
is used, the uncertainty in varying the scale from $Q^2/2$ to $2Q^2$ is 
sufficient to give agreement with the data. Given this theoretical uncertainty,
it is clear that no firm conclusion can be drawn on the presence of non-GLAP
evolution in forward jet production.  
\subsection{Forward $\pi^0$ production at HERA}
\label{sec:forwardpi0}
Since the size of any BFKL effect will certainly strongly increase as $x$ falls,
processes that can reach lower values of $x$ than possible for forward jets
could be very valuable. In addition, the use of a single 
forward-going particle as a probe 
reduces the uncertainty due to jet finding algorithms as well as lowering the
minimum angle which can be probed, since single particle shower profiles are
significantly narrower than comparable energy jets. Such considerations led the
H1 Collaboration to investigate the production of forward $\pi^0$s. The use
of very energetic $\pi$s permits the correspondence between leading particles
and the struck parton to be used to isolate a region in which BFKL effects could
be important. In principle any particle species could be used; however, the 
power of the central tracking systems used in the HERA detectors is weakest in
this region and the calorimetry in general permits both a reasonable 
identification
of $\pi^0$s as well as sensitivity to smaller angles.  

The H1 collaboration has isolated~\cite{pl:b462:440} a $\pi^0$ signal from 5.8 
pb$^{-1}$ of data taken in 1996. The $\pi^0$ candidates were required to have a 
transverse momentum in the hadronic CMS of greater than  2.5~GeV and a $Q^2$ 
range $2 < Q^2 < 70$ GeV$^2$.
They were isolated
between polar angles of $5^{\circ}$ and $25^{\circ}$ and required to have energy 
$x_\pi > 0.01 \cdot E_p$, where $E_p$ is the proton beam energy of 
820 GeV. At such high energies, the two photons from the $\pi^0$
decay cannot be separated. Instead, they are identified by a detailed 
analysis of the longitudinal and transverse shape
of the energy depositions in order to separate electromagnetic from
hadronic showers. About 600 $\pi^0$ candidates were found with $p_T >
3.5$ GeV. The efficiency for detection was around 45\%. 

\begin{figure}[h]
\begin{center}
\epsfig{file=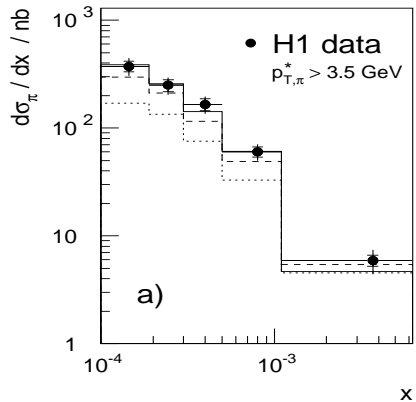,%
      width=6cm,%
      height=6cm%
        }
\end{center}
\caption{The differential cross section as a function
of $x$ for forward $\pi^0$s from the H1 Collaboration
for $2 < Q^2 < 70$ GeV $^2$.
The $\pi^0$s were required to have $p_T > 3.5$ GeV in the hadronic
centre-of-mass system, $x_\pi > 0.01$ and angle with respect
to the proton direction between $5^o$ and $25^o$.
The dashed line shows the prediction from RAPGAP and the dotted
line that from LEPTO. The solid line shows the result of a modified
LO BFKL calculation (see text).}
\label{fig:H1:pizerosigma}
\end{figure}

Figure~\ref{fig:H1:pizerosigma} shows the differential cross-section $d\sigma/dx$
for $p_T(\pi^0) > 3.5$ GeV. 
Also shown are predictions from the
RAPGAP and LEPTO Monte Carlo programs, as well as the prediction from a
modified LO BFKL calculation~\cite{epj:c9:611} convoluted with $\pi^0$ 
fragmentation functions. The LEPTO model~\cite{cpc:101:108} 
does not give a good
description of the data. A considerable improvement is given by a model, 
RAPGAP2.06~\cite{cpc:86:147},
which includes resolved virtual photons in the hard scattering process. Such
resolved processes have been shown to be important in DIS in some kinematic 
regions~\cite{pl:b479:37} even at moderate $Q^2$. Nevertheless, even 
taking into
account the uncertainty caused by varying the renormalisation scale, RAPGAP 
cannot fit the data over the full kinematic range.  However, the ARIADNE model,
which is not shown in the figure, can give a good description of the data, 
although there is considerable arbitrariness in its predictions.  
The LO BFKL parton calculation is in good
agreement with the data. However, once again there is a large uncertainty
caused by a variation in the renormalisation scale of a factor two above and
below the nominal value, leading to a 60\% variation in the prediction. 

Although the agreement of the BFKL model with the data is interesting, overall
the inherent uncertainties in the various models are such that it is difficult
to draw any clear conclusion as to the presence of BFKL effects in the data.
\subsection{BFKL tests at the Tevatron}
\label{sec:BFKL:Tev}
The Tevatron gives access to a rather different kinematic range in which BFKL
effects could possibly become important. Here, in the production of high-energy 
jets, the centre-of-mass energy can be much larger than the momentum transfer, 
$Q$, so that the jet cross section contains large logarithms, $\ln (s/Q^2)$,
which must be summed to all orders. Such a summation can be achieved using the BFKL 
formalism. The D0 Collaboration\cite{prl:84:5722} has isolated dijet events with very large
rapidity separations and measured the cross section as a function of $x_1$, 
$x_2$ and $Q^2$, where 1 and 2 label the most-forward and most-backward jets,
respectively. The longitudinal momentum fractions
of the proton and antiproton, $x_1$ and $x_2$,
carried by the two interacting partons are defined as:
\begin{equation}
x_{1,2} = \frac{2 E_{T_{1,2}} }{\sqrt{s}} \; e^{ \pm \overline{\eta}} \;
                                        \cosh(\Delta \eta /2) \;
\label{eq:x12}
\end{equation}
where $E_{T_1}(E_{T_2})$ and $\eta_1(\eta_2$) are the 
transverse energy and pseudorapidity of the most forward(backward) jet,
$\Delta \eta = \eta_1 - \eta_2 \geq 0$,
and $\overline{\eta}=(\eta_1+\eta_2)/2$. Thus, by measuring the cross section
at the largest accessible values of $\Delta \eta$, the separation in $x$ of the
colliding partons is maximised, thereby optimising the phase space for
gluon ladders with strong ordering in $x$, i.e. BFKL evolution, to be
produced. The momentum transfer during the hard scattering is defined as:
\begin{equation}
Q = \sqrt{E_{T_1}E_{T_2}} 
\nonumber
\end{equation}
The BFKL prediction~\cite{np:b282:727} for the cross 
section is:
\begin{equation}
\hat{\sigma}_{\rm BFKL} \propto
                  \frac{1}{Q^2} \cdot
                  \frac{e^{(\alpha_{BFKL}-1)\Delta \eta}}
{\sqrt{\alpha_s \Delta \eta}} \;
\label{eq:sigma_bfkl}
\end{equation}
where $\alpha_{BFKL}$ is the BFKL intercept that governs the strength of the 
growth of the gluon distribution at small $x$, which 
in leading logarithmic approximation (LLA), is given 
by:
\begin{equation}
\alpha_{BFKL} - 1 = \frac{\alpha_S(Q) \, 12 \ln2 }{\pi}
\label{eq:ap}
\end{equation}
\begin{figure}[h]
\begin{center}
\epsfig{file=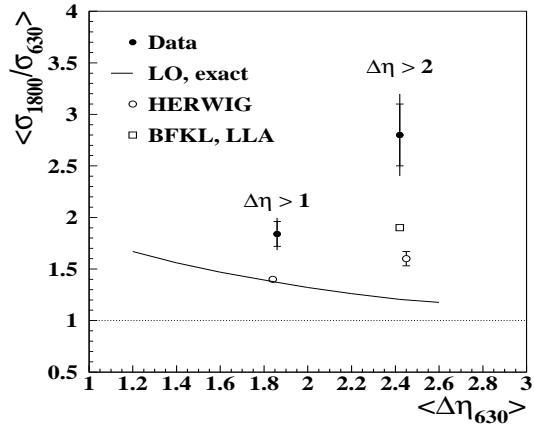,%
      width=8cm,%
      height=6cm%
        }
\end{center}
\caption{The ratio of jet cross sections with rapidity gaps between them measured at $\surd s = $ 1800 and 630 GeV by the D0 Collaboration. The 
data points are plotted as filled circles 
for rapidity gaps greater than 1, and 2,  
as a function of the mean rapidity gap at 630 GeV. The light shaded circles 
represent the predictions from HERWIG while the square represents a calculation
based on the LLA BFKL approach. The thick error bars denote the statistical
uncertainty, while the thin error bar denotes the statistic and systematic
uncertainties added in quadrature.}
\label{fig:D0:BFKL}
\end{figure}

The cross section is a convolution of the probability to find the
interacting partons with given values of $x$ inside the proton, together
with the partonic scattering cross sections that contain the BFKL
effects of interest. 
In order to remove the dependence of the cross section on the parton
distribution functions, it is convenient to measure the cross sections
at different centre-of-mass energy and take the ratio at fixed $x_1, x_2$
and $Q^2$; this also reduces some correlated systematic effects 
stemming from the detector characteristics. The ratio of cross sections
can be obtained from equation~\ref{eq:sigma_bfkl}:
\begin{equation}
R \equiv
\frac{\sigma(\sqrt{s_A})}
         {\sigma(\sqrt{s_B})}
= \frac{\hat{\sigma}(\Delta \eta_A)}
         {\hat{\sigma}(\Delta \eta_B)}
  = \frac{e^{(\alpha_{BFKL}-1)(\Delta \eta_A-\Delta \eta_B)}}
{\sqrt{\Delta \eta_A/\Delta \eta_B}} 
\label{eq:D0:BFKL:ratio}
\end{equation}
Clearly, the greater the difference between the two centre-of-mass energies, 
the greater the effective variation in $\Delta \eta$ between the
two data sets and thereby the
larger the BFKL effects should become. 

The data samples used in the D0 analysis was taken at $\surd s =$ 1800 and 630 
GeV. The $x_1, x_2$ ranges were restricted to lie between 0.06 and 0.22 in order 
to avoid detector biases. Only one $Q^2$ bin was used, with $400 < Q^2 < 1000$
GeV$^2$. Figure~\ref{fig:D0:BFKL} shows the ratio of the cross sections at the
two centre-of-mass energies as a function of the mean separation in $\eta$ in 
the 630 GeV data set. 
Also shown on the plot are the predictions from LO GLAP
evolution, from the HERWIG Monte Carlo, and from the BFKL calculation. The data
agree with none of the models, showing a much steeper increase than predicted 
even by the BFKL calculation. However, the steep rise is certainly more in
accord with the BFKL prediction than that of LO GLAP evolution; 
what the rise in the HERWIG predictions means is unclear, at least to this
author. In conclusion, the situation is yet again confused.

To summarise this section, despite having examined specific exclusive processes in which the effects of
BFKL evolution are expected to be maximal, the current experimental
situation is that there
is little firmer evidence for deviations from standard GLAP evolution than
there was in the inclusive $F_2$ data.
\section{Interpretation and Models}
\label{sec:interpretation}
While the standard perturbative QCD GLAP evolution reigns supreme at medium $x$
and high $Q^2$, the area of low $x$ is a particularly rich and complex area in 
which this perturbative QCD {\it anzatz} meets and competes with a large variety 
of other approaches, some based on QCD, others either on older paradigms such as 
Regge theory or essentially {\it ad-hoc} phenomenological models. In this 
section a far-from-exhaustive survey of such models is undertaken. Often it is
convenient to concentrate on a particular model in order to confront specific
predictions with the data. This in no way necessarily implies that such an 
example model is the best one available, or that others can be ignored in
its favour - rather it is an attempt to illustrate generic characteristics
of particular classes of approach to the elucidation of the data.

In this section the older, Regge-based approach to the understanding of $F_2$ is 
examined, followed by models that give simple parameterizations of the structure 
functions based, more or less loosely, on pQCD. A brief overview of the current 
state of global fits to the parton distribution functions is then carried out,
followed by a discussion of new data from ZEUS on the logarithmic derivatives
of $F_2$ and possible implications for QCD evolution. Dipole models, in 
particular that of Golec-Biernat \&
W\"{u}sthoff~\cite{pr:d59:014017,pr:d60:114023} are discussed in some detail
and compared to the data.
\subsection{Regge-based models of $F_2$}
\label{sec-Regge}
Regge theory~\cite{ncim:14:951,ncim:18:947,collins:1977:regge} has a long and 
distinguished history in
particle physics as a conceptual basis linking bound-state spectra, forces
between particles and the cross-section behaviour as a function of energy over a 
wide variety of processes through the analytical properties of high-energy 
scattering amplitudes. The basis of the theory is that solutions of the 
Schr\"{o}dinger equation for non-relativistic potential scattering can be
solved in terms of a complex angular momentum variable, $j$. 
For many potentials,
the only singularities of the scattering amplitude are poles
in the complex angular momentum plane, known as 'Regge poles'. Somewhat
surprisingly, such non-relativistic concepts have proven to be very useful
in particle physics. To take the specific example relevant to this discussion,
any total cross section having a power-law dependence on the centre-of-mass 
energy lends itself to a simple explanation in terms of Regge poles 
corresponding to the exchange of specific particles. The most important poles 
are those corresponding to the exchange of vector and tensor mesons, which
give a pole at $j$ = 1/2, and that near to $j=1$, which corresponds to a 
particle with the quantum numbers of the vacuum that has never been
observed as a final-state particle. This particle is known as the Pomeron,
and is thought of as the exchanged particle that mediates elastic and
diffractive scattering. In fact, if the pole is assumed to be at $j=1.08$,
known as the `soft Pomeron' singularity, then it gives a good description
of the cross sections of a wide variety 
of hadronic processes~\cite{np:b231:189}.

\begin{figure}[h]
\begin{center}
\epsfig{file=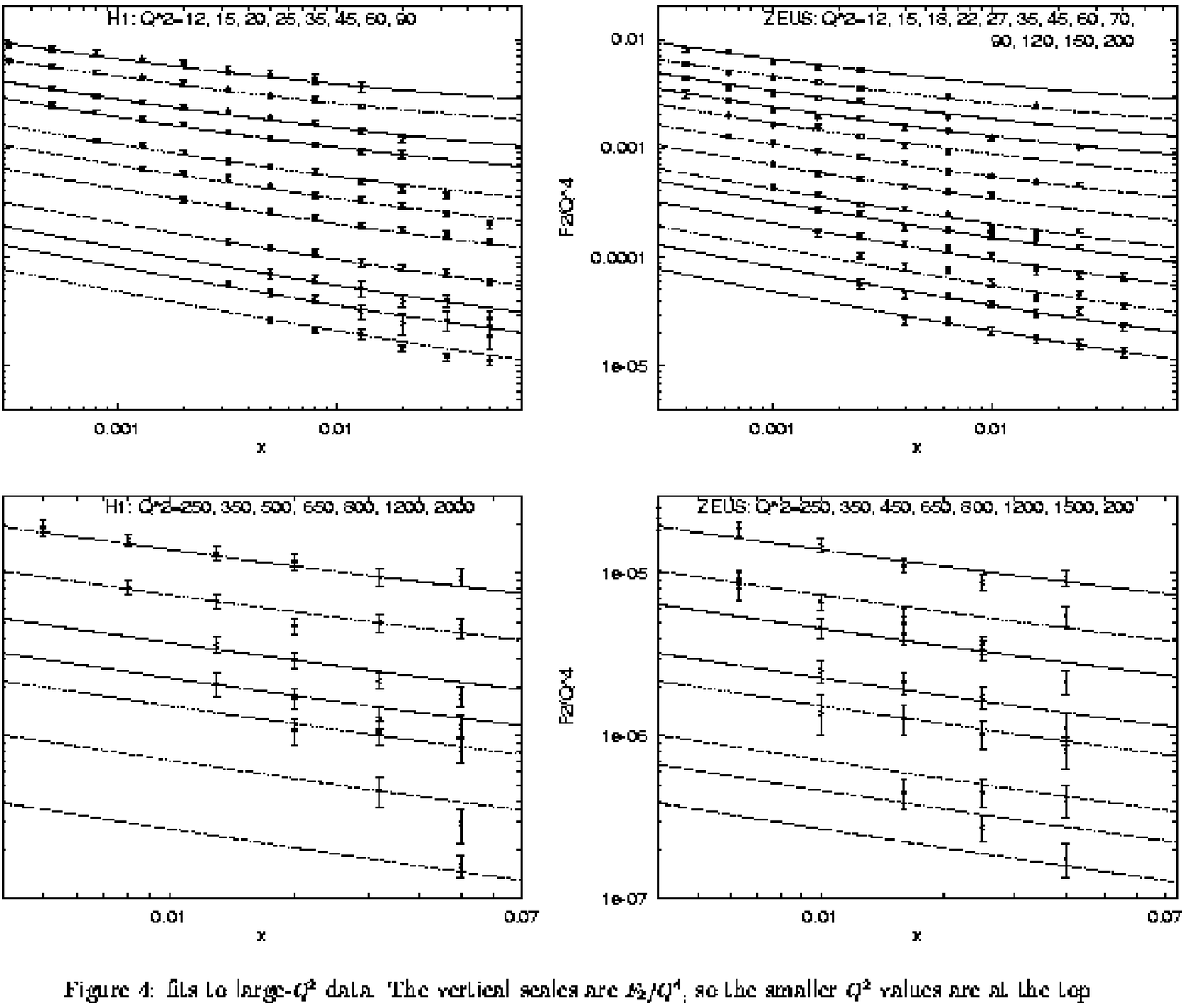,%
      width=6cm,%
      height=6cm,%
      clip=%
        }
\end{center}
\caption{The H1 published data on $F_2$ plotted divided by
$Q^4$ in different $Q^2$ bins as a function of $x$. The lines
show Donnachie and Landshoff's fits using Reggeon, soft and 
hard Pomeron contributions.}
\label{fig:DL:f2}
\end{figure}

The advent of HERA, allowing access to large $W^2$ in both soft and hard
processes, offered a fertile ground for the
application of Regge theory, since it works {\it par excellence} at very high 
energies; for example, in DIS when $W^2$ is much greater than any other
invariant. Indeed, equations~\ref{eq:w2q2x} and ~\ref{eq:f2siglt} make it
clear that, at low $x$, $F_2$ can be expressed in terms of high-energy 
transverse and longitudinal cross sections. It was therefore interesting to
discover that the strong rise in $F_2$ at low $x$ discovered at HERA,
which can be parameterised
as $x^{-\lambda} \sim W^\lambda$, was incompatible with the soft Pomeron 
intercept that had worked so well in many other reactions. Donnachie and
Landshoff~\cite{pl:b437:408} 
suggested that a good description could be achieved if a further
simple Regge pole were assumed at $j = 1.435$, the so-called `hard Pomeron'. 
They fit to the form:
\begin{equation}
F_2(x,Q^2) = \sum_{i=0}^3 f_i(Q^2) x^{-\epsilon_i}
\label{eq:DL:hP}
\end{equation}
and, as can be seen from figure~\ref{fig:DL:f2}, such an {\it ansatz}
does indeed
give a good description of the $F_2$ data at low $x$ (and indeed at higher
$x$ as well). 

\begin{figure}[h]
\begin{center}
\epsfig{file=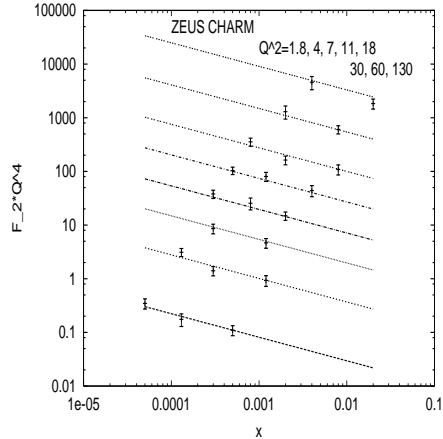,%
      width=6cm,%
      height=6cm%
        }
\end{center}
\caption{ZEUS data on $ F_2^{c\overline{c}}$ plotted divided by $Q^4$
as a function of $x$. The lines show the Donnachie-Landshoff prediction
with only the hard Pomeron term in the fit.}
\label{fig:DL:charm}
\end{figure}

Donnachie and Landshoff also looked at the ZEUS data on the charm
structure function, $F_2^{c\overline{c}}$~\cite{pl:b470:243}. Here, presumably
because the mass of the charm quark provides a hard scale, there is no
requirement for the vector and tensor meson or the `soft Pomeron'
pole, leaving only one term in equation~\ref{eq:DL:hP}. The fit shown in 
figure~\ref{fig:DL:charm} to the `hard Pomeron'
term only gives a good fit to the data.
 
Despite the successes of this approach, there are several major drawbacks.
Firstly, the $f_i(Q^2)$ in equation~\ref{eq:DL:hP} cannot be predicted from
Regge theory. Having added one extra Pomeron, 
there is no logical reason not to add others when HERA data
from other channels~\cite{epj:c14:213,pl:b478:146,misc:bruni:dis2000} 
and improved accuracy imply deviations from a `universal' two-Pomeron {\it 
ansatz}.
Indeed, more complex singularities than poles can also be 
added~\cite{pl:b309:191,pl:b459:265,pr:d61:034019} leading
to behaviour more complex than simple powers of $W$.   
Thus, the Regge description can have many essentially
arbitrary parameters and risks becoming merely a phenomenological 
parameterisation. The great range of seemingly disparate
phenomena that can be explained in the Regge framework nevertheless
implies that any more comprehensive theory of the strong interaction,
{\it viz.} QCD, must somehow `explain' or assimilate Regge concepts in
a natural way. This line of investigation is a fruitful one currently
being actively pursued and also leads naturally into the next section.
\subsection{QCD-inspired parameterisations}
\label{sec:QCDinsp}
The double-logarithmic limit of QCD, in which both $Q^2$ and $1/x$ become
very large, has long been known to imply~\cite{pr:d10:1649} that $F_2$ should fit 
to the form:    
\begin{equation}
F_2(x,Q^2) \propto \exp{\{\sqrt{(48/\beta_0) \ln (1/x) \ln \ln Q^2} \}}
\label{eq:doublelnscale}
\end{equation}
where $\beta_0$ is the standard Renormalisation Group Equation $\beta$ function. 
Ball and Forte~\cite{pl:b335:77,pl:b336:77} showed that indeed the HERA data
were well represented by this form, which they refer to as `double-asymptotic
scaling'.

\begin{figure}[h]
\begin{center}
\epsfig{file=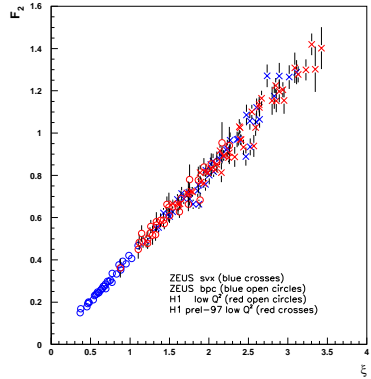,%
      width=6cm,%
      height=6cm%
        }
\end{center}
\caption{ZEUS and H1 $F_2$ data plotted against the scaling variable
$\xi$ (see equation~\protect\ref{eq:F2:Haidt}).}
\label{fig:Haidt}
\end{figure}

Other authors have examined functional forms that are related to the
double-asymptotic scaling of equation~\ref{eq:doublelnscale}.  
Buchm\"{u}ller and Haidt~\cite{hep-ph-9605428} set out the theoretical region
of validity of the so-called `double-logarithmic' scaling regime. 
The functional form\footnote{Modified~\cite{proc:dis:1997:386} from the original 
form~\cite{proc:dis:1996:179} proportional to $\ln Q^2/Q_0^2$ in order to take
account of new data at very low $Q^2$.} is:
\begin{equation}
F_2(x,Q^2) = m \ln\left(\frac{Q^2+Q_0^2}{Q_0^2}\right)\ln\left(\frac{x_0}{x}
\right) = m \xi
\label{eq:F2:Haidt}
\end{equation}
where $m$ is a constant to be determined from the data. Figure~\ref{fig:Haidt}
shows~\cite{proc:dis:1999:186} HERA data\footnote{The data at
the lowest $Q^2$  was still preliminary at the time of this conference.} plotted as a function of $\xi$ to this functional form.
A linear fit~\cite{proc:dis:1999:186} is remarkably good. 
This functional form is appropriate to the Regge limit of fixed $Q^2$ and 
$x \rightarrow 0$, and indeed it corresponds to the first term of the
more general solution to the double-asymptotic form {\it viz.} a summation
of all terms of the form~\cite{hep-ph-9605428}
\begin{equation}
\left(\alpha_s \ln \frac{Q^2}{\Lambda^2} \ln \frac{1}{x}\right)^n
\label{doublelogterm}
\end{equation}
which rises faster than any power of $\ln 1/x$ as $x$ falls. In the
more general case in which both $Q^2$ and $1/x$ become large, or at sufficiently 
small $x$, the double-asymptotic form ought to become more appropriate. In fact, 
somewhat surprisingly, it would seem that even at very small $x$ the 
double-logarithmic form gives a good fit to the data.

\begin{figure}[h]
\begin{center}
\epsfig{file=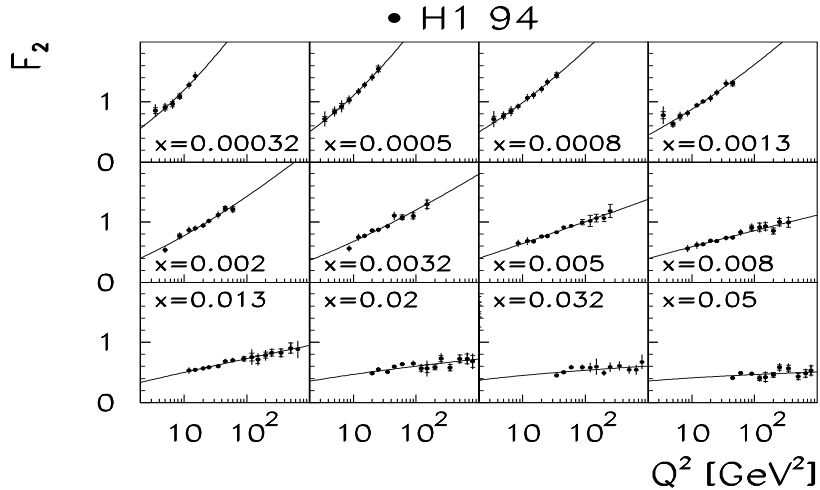,%
      width=12cm,%
      height=8cm%
        }
\end{center}
\caption{Preliminary H1 $F_2$ data fit to the form of 
equation~\protect\ref{eq:erdmann}.}
\label{fig:erdmann:H1fit}
\end{figure}

A parameterisation clearly related to both the double-asymptotic and 
double-logarithmic ones has recently been used by
Erdmann~\cite{misc:erdmann:dis2000}, in particular
to facilitate comparison between the proton, photon and Pomeron structure functions. It has the form:
\begin{equation}
F_2(x,Q^2) = a(x) \left[ \ln \left(\frac{Q^2}{\Lambda^2}\right)
\right]^{\kappa(x)} 
\label{eq:erdmann}
\end{equation}
where $\Lambda$ is the QCD scale parameter. 
Figure~\ref{fig:erdmann:H1fit} shows that this form gives a good fit to the
preliminary H1 $F_2$ data. 

It is also true for the published ZEUS and the BCDMS
data, even at high $x$. In some sense equation~\ref{eq:erdmann} represents
the alternative approximation to the Regge limit of Haidt, i.e. fixed $x$
and $Q^2$ large. The fact that both $a$ and $\kappa$ are functions of $x$
however allows the $\ln 1/x$-like terms in the sum of
equation~\ref{doublelogterm} to be
approximately taken into account, so that the parameterisation also
works well at low $Q^2$ and $x$. It can be seen by
comparison with equation~\ref{eq:F2:MSbar} that
$a(x)$ is proportional to the charged parton distribution at a
particular value of $Q^2$, {\it viz.} that for which $\ln Q^2/\Lambda^2$ = 1,
and that $\kappa$ is related to the scale-breaking of $F_2$ and therefore
to gluon radiation from the partons. 

\begin{figure}[h]
\begin{center}
\epsfig{file=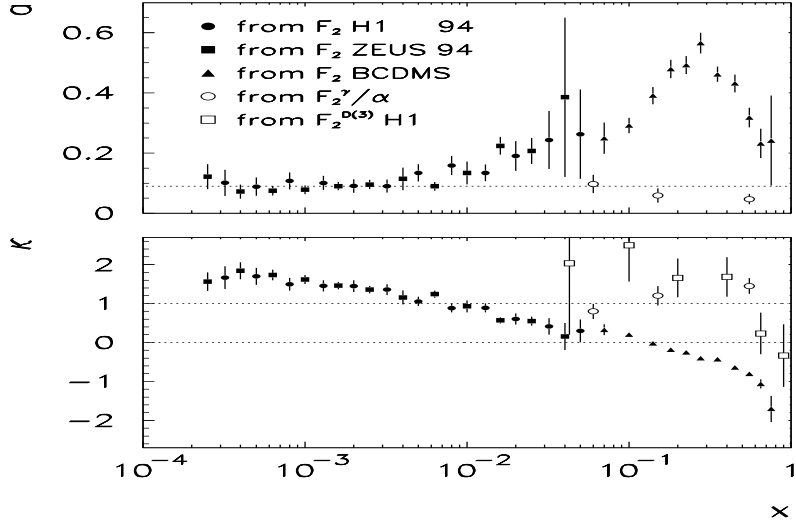,%
      width=12cm,%
      height=8cm%
        }
\end{center}
\caption{The values of the coefficients
in equation~\protect\ref{eq:erdmann} as a function of $x$
after fits of this functional
form to BCDMS, H1 and ZEUS $F_2$
data (filled symbols). The open symbols correspond to photoproduction
and diffractive data that are not discussed in the text.}
\label{fig:erdmann:akappa}
\end{figure}

Figure~\ref{fig:erdmann:akappa} shows the values of $a$ and $\kappa$ as
a function of $x$. The valence quark distribution at high $x$ can be
very clearly seen, as well as that fact that for the low-$x$ regime of
interest here, $a(x)$ is approximately constant. This implies that
the scale-breaking represented by $\kappa$ is what drives the
increase in $F_2$. Figure~\ref{fig:erdmann:akappa} shows that $\kappa$ increases
more or less linearly from the negative scale-breaking at high $x$ until
at least $x \gsim 10^{-4}$, at which point it seems to level off. 
The implications of this observation will be discussed in more detail
in section~\ref{sec:sat} in conjunction with the ZEUS data at very low $x$.
\subsection{Global fits}
\label{sec:global}
The extension of the kinematic range and the high-precision data on $F_2$
from HERA provided a substantial impetus to the determination of parton
distribution functions via global fits to a wide variety of data. The major
approaches are due to the CTEQ group~\cite{epj:c12:375}, Gl\"{u}ck, Reya and Vogt (GRV)~\cite{epj:c5:461} and Martin et al.
(MRST)~\cite{proc:dis99:1999:105}. In general
all groups fit to data from fixed target muon and neutrino deep inelastic
scattering data, the HERA DIS data from HERMES, H1 and ZEUS, the
$W$-asymmetry data from the Tevatron as well as to
selected process varying from group to group such as prompt photon data from
Fermilab as well as high-$E_T$ jet production at the Tevatron. The different data sets give different sensitivity to the proton distributions depending
on the kinematic range, but together constrain them across almost the whole
kinematic plane, with the possible exception of the very largest values of $x$,
where significant uncertainties still remain~\cite{epj:c13:241}. 

\begin{figure}[h]
\begin{center}
\epsfig{file=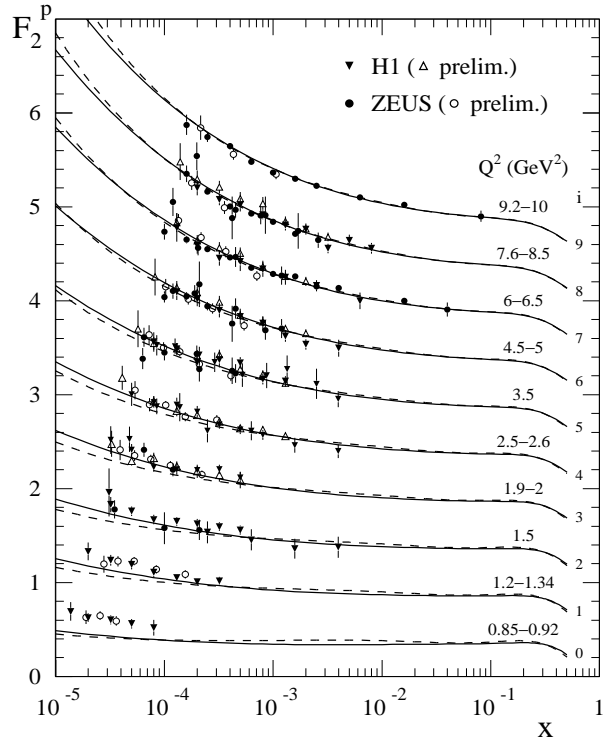,%
      width=8cm,%
      height=10cm%
        }
\end{center}
\caption{The GRV98 fit to the ZEUS and H1 $F_2$ data in the
low-$Q^2$ region in bins of $Q^2$
plotted as a function of $x$.}
\label{fig:GRV98}
\end{figure}

The approach of GRV is somewhat different from that of the other two
groups. They utilise the fact that, as $Q^2$
$\rightarrow 0$, parton distributions are fully constrained by the charge  
and momentum sum rules. By assuming valence-like distributions for the  
quarks at a very low starting $Q^2$, in principle the gluon and sea  
distributions can be generated purely dynamically. However, it is 
found that such a procedure generates parton distributions which are too  
steep as $x$ decreases. Instead they 
input `valence-like' distributions for both quarks and  
gluons fixed by high-$x$ data at a larger though still very small $Q^2$. 
The starting value, $Q^2_0$, is determined by the point at which  
the input gluon distribution is of the same order as the input $u$ valence  
quark distribution and is $\sim 0.5$ GeV$^2$ in NLO  
QCD~\cite{zfp:c53:127}. Although there are quite large uncertainties on the  
value of $Q^2_0$ and on the valence-like distributions assumed at  
$Q^2_0$, the effect of these is suppressed in the comparison with the
high-$Q^2$ HERA data by the long evolution distance. In general, the GRV
parameterisation gives good fits to the HERA data,
as shown in figure~\ref{fig:GRV98}, although as $Q^2 \rightarrow Q^2_0$
the fit becomes worse, as would be expected from the formalism. In addition,
however, GRV at NLO has difficulties in fitting the logarithmic derivatives
of $F_2$ for values of $x < 10^{-3}$~\cite{rmp:71:1275} (although see
section~\ref{sec:sat}).

\begin{figure}[h]
\begin{center}
\epsfig{file=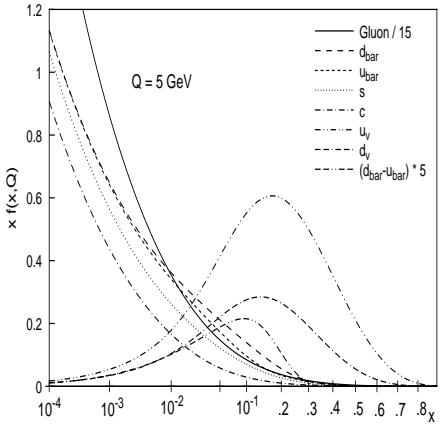,%
      width=6cm,%
      height=6cm%
        }
\end{center}
\caption{The PDFs resulting from the CTEQ5M fits at $Q^2$ = 25 GeV$^2$.}
\label{fig:CTEQ:PDFs}
\end{figure}

The approaches of CTEQ and MRST are basically similar, although they differ both
in the data sets used as well as in the fitting procedure and the technical details of the theoretical tools used, e.g. the treatment of heavy quarks in DIS. In their latest fits CTEQ prefer to omit the prompt photon data because
of the uncertainties in scale dependence and the appropriate value for the
intrinsic $k_T$ required to fit the data. Instead they use single-jet inclusive
$E_T$ distributions to constrain the gluon distribution at large $x$.
In contrast, until their most recent publication, MRST retained
the prompt photon data, giving alternative PDFs
depending on the value for the prompt-photon intrinsic $k_T$ used. Both
groups parameterise the parton distributions
in terms of powers of $x$ and $(1-x)$ leading to fits with many
free parameters. The MRST NLO parameterisation of the gluon is
shown below as an example:
\begin{eqnarray}
xg &=& A_{g}x^{-\lambda_g} (1-x)^{\eta_g}
(1+\epsilon_{g}\sqrt{x}+\gamma_{g}x)
\label{eq:MRST:NLOg}
\end{eqnarray}
where $A_{g}, \lambda_g, \eta_g, \epsilon_g$ and $\gamma_g$
are free parameters in the fit.
The treatment of the $d/u$ ratio at high $x$ has recently been addressed
by Yang and Bodek~\cite{epj:c13:241}, who point out that deuterium binding corrections should
be applied to the NMC $F_2^n/F_2^p$ data. Such corrections give good
fits in the global analyses, except to the uncorrected NMC data 
themselves.  
The PDFs determined 
from the CTEQ5M fit are shown in figure~\ref{fig:CTEQ:PDFs} at $Q^2 =$
25 GeV$^2$.
The steep rise of the gluon and sea distributions as $x$ falls is evident.

\begin{figure}[h]
\begin{center}
\epsfig{file=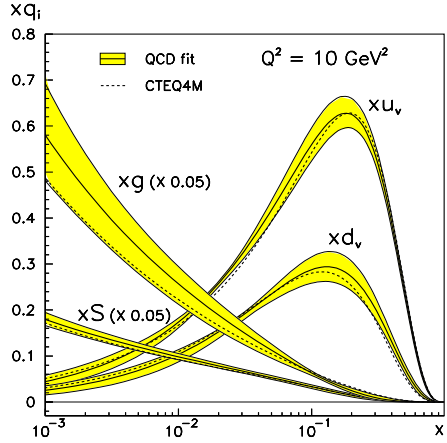,%
      width=6cm,%
      height=6cm%
        }
\end{center}
\caption{The valence quark, sea quark and gluon PDFs resulting
from the Botje fit. The bands show the
uncertainty associated with each PDF.}
\label{fig:Botje:PDFs}
\end{figure}

A long-standing problem with the various global PDFs has been the fact
that no error was associated with the central values. The difficulties
associated with producing such errors from a multi-parameter
fit to many data sets with differing correlations are certainly
formidable. It is therefore an extremely important and welcome development
that Botje has recently produced for the first time PDFs with associated
error matrices~\cite{epj:c14:285}. Figure~\ref{fig:Botje:PDFs} shows the
valence- and sea-quark distributions together with that of the gluon
from Botje's analysis. 

\begin{figure}[h]
\begin{center}
\epsfig{file=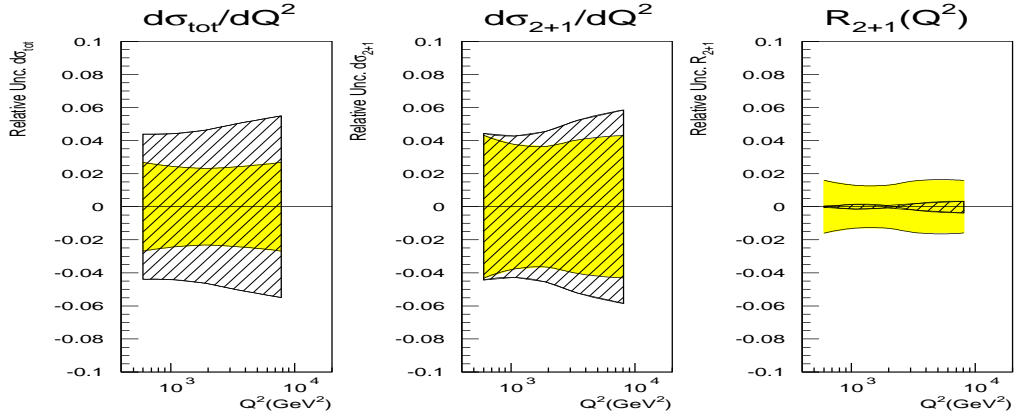,%
      width=15cm,%
      height=6cm%
        }
\end{center}
\caption{The uncertainty of three quantities used in the ZEUS determination
of $\alpha_s$ from dijets in DIS. The cross section differential in $Q^2$, the dijet differential cross section, and the ratio
of the dijet to the total cross section are shown. The shaded area
shows the effect of taking the correlated errors produced
by the Botje fit properly into account,
the hatched area that of ignoring the correlations.}
\label{fig:Botje:2+1jet}
\end{figure}

The fit utilises a more restricted range of data than
the CTEQ and MRST fits, using the H1 and ZEUS $F_2$ data together with the fixed
target muon and neutrino data; the Drell-Yan data from E866~\cite{prl:80:3715} are used to constrain the $\overline{u} - \overline{d}$ distribution.
Despite the more restricted data sets used, the results of the fit are
very compatible with the most recent fits of CTEQ and MRST. The importance
of the error matrices produced by this fit can be illustrated by the
example shown in figure~\ref{fig:Botje:2+1jet}. 
Here the effect of the
errors on the uncertainty in the prediction of various
cross sections and ratios in the ZEUS DIS dijet analysis~\cite{misc:Osaka:891}
is shown. The very large difference in the estimated error, depending on whether the correlations between the parameters in the PDFs are taken into account
of not, is striking and of the highest importance in a realistic calculation of the error on quantities such as $\alpha_s$.          

\begin{figure}[h]
\begin{center}
\epsfig{file=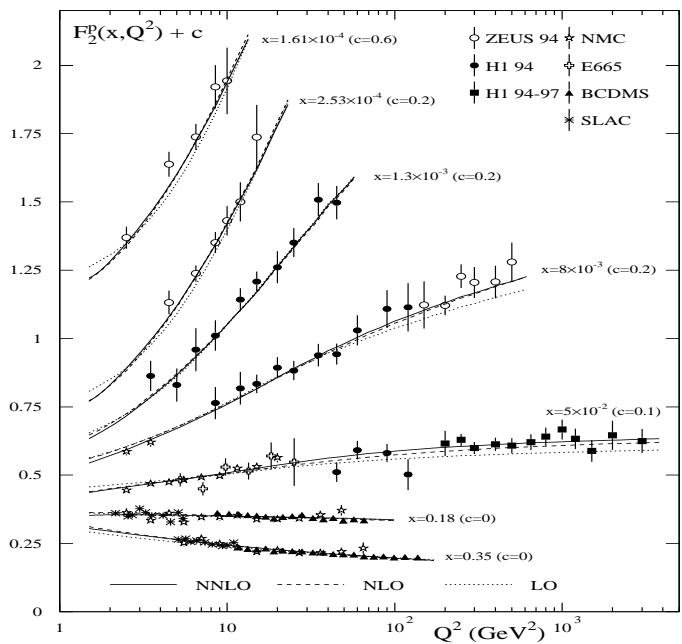,%
      width=10cm,%
      height=10cm%
        }
\end{center}
\caption{The MRST `central' NNLO fit to DIS data. The solid line
shows the NNLO fit, while the NLO fit is shown by the dashed line
and the LO fit by the dotted line. The data are from H1, ZEUS
and the fixed target experiments and are plotted 
in $x$ bins as a function of $Q^2$ with an additive constant
added to the data of each $x$ bin to improve visibility.}
\label{fig:MRST:NNLOfit}
\end{figure}

All of the parameterisations discussed above were carried out in the framework
of NLO QCD. With the increasing precision of the DIS data, as well as the
need for accurate predictions of cross sections at the LHC, the need for 
next-to-next-to-leading-order (NNLO) fits is obvious. The first steps in
this regard have already begun, and some moments of the NNLO splitting functions
have already been calculated~\cite{np:b492:338}. Using this with other
available information, van Neerven and Vogt~\cite{np:b568:263,hep-ph-0006154} have produced analytical expressions for the splitting functions which represent
the slowest and faster evolution consistent with the currently available
information. The MRST group has recently used this information to investigate
NNLO fits to the available data~\cite{hep-ph-0007099}. Such an analysis requires
some changes to the parameterisations used, so that for example the NLO
parameterisation of the gluon of equation~\ref{eq:MRST:NLOg} becomes:
\begin{equation}
 xg (x, Q_0^2) \; = \; A_g \: x^{-\lambda_g} \: (1 - x)^{\eta_g} \:  
(1 + \varepsilon_g \sqrt{x} +  
\gamma_g x) \: - \: A_g^\prime \: x^{- \lambda_g^\prime} \:  
(1 - x)^{\eta_g^\prime} 
\label{eq:MRST:NNLOg}
\end{equation}
primarily in order to facilitate a negative gluon density at low $x$ and
low $Q^2$, which, although conceptually somewhat bizarre, is nevertheless preferred by the fits, even at NLO. The results of the `central' fit, between the extremes of the van~Neerven-Vogt parameterisation, is shown in figure~\ref{fig:MRST:NNLOfit}.

\begin{figure}[h]
\begin{center}
\epsfig{file=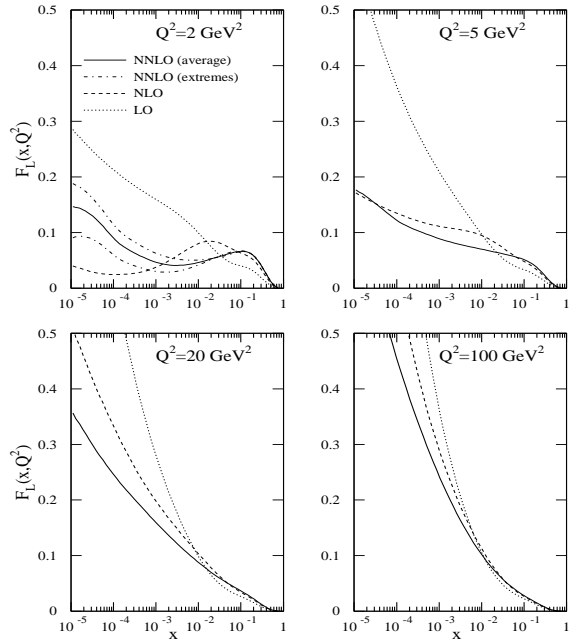,%
      width=8cm,%
      height=10cm%
        }
\end{center}
\caption{The $F_L$ structure function from the MRST fits, taking into account
part of the NNLO corrections in four bins of $Q^2$ as a function of $x$.
The solid line shows the 'average' of the parameter space available
to choose the NNLO parameters, while the dashed-dotted lines show the two extreme possibilities. The NLO fit is indicated by the dashed 
line while the LO fit is indicated by the dotted line.}
\label{fig:MRST:NNLOPDFs}
\end{figure}

There are also changes of the LO and NLO fits with respect to earlier
publications, in as much as MRST now follow CTEQ in using the Tevatron
high-$E_T$ data rather than the prompt-photon data, and preliminary HERA
$F_2$ data has been included in the fit. There is a marked improvement
in the quality of the fit in the progression LO $\rightarrow$ NLO
$\rightarrow$ NNLO, in particular in terms of the NMC data. The size
of higher-twist contributions at low $x$ also decreases, so that at
NNLO is it essentially negligible. The effect of going to NNLO on the
PDFs themselves is highly non-trivial. This is illustrated in
figure~\ref{fig:MRST:NNLOPDFs}, where the quite major changes in 
$F_L$, particularly at low $x$, are evident. 
There is also a large variation depending on the choices made
in the parameter space allowed by the partial NNLO
{\it ansatz}.  Indeed, the
GLAP approach is not convergent for $Q^2 < 5 $GeV$^2$, which may well
be due to the neglect of important $\ln 1/x$ contributions. 
However, the instability seen at low $Q^2$ soon
vanishes at higher $Q^2$.

\begin{figure}[h]
\begin{center}
\epsfig{file=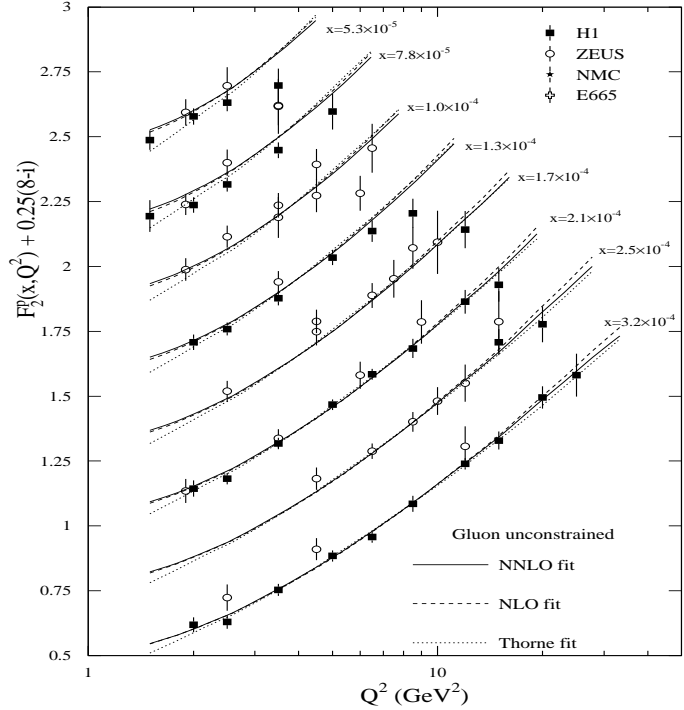,%
      width=9cm,%
      height=10cm%
        }
\end{center}
\caption{The NNLO MRST fit, modified by Thorne to include a NLO
BFKL kernel. The `standard' NNLO fit is shown as the full line, 
the dashed line shows the NLO fit and the dotted line the Thorne
BFKL modification.}
\label{fig:Thorne:BFKL}
\end{figure}

Thorne has indeed investigated the question of incorporating $\ln 1/x$
terms in the splitting functions by incorporating the solution of the 
NLO BFKL kernel using a running coupling 
constant~\cite{pl:b474:373,*misc:DIS2000:Thorne}.
The results are shown in figure~\ref{fig:Thorne:BFKL}. 
It is clear that
the inclusion of the BFKL terms does indeed give an improved fit compared to
the `central' NNLO fit, particularly at the lowest $Q^2$ and $x$. This
may be one of the first unambiguous 
indications of the importance of BFKL evolution;
if so, it is rather surprising that it has occurred in the analysis of the
inclusive data, rather than the exclusive channels, expected to be more sensitive, that were examined in section~\ref{sec:otherprobes}. 

In conclusion, there have been major advances in the field of parton 
distribution functions and global fits in the last year. Not only is
there now a parameterisation which gives produces associated
error matrices, which
is of first importance in the treatment and extraction of experimental
results, but also the first attempts to incorporate NNLO corrections into the
fitting has begun. In the latter case, it is clear that there is still
a great deal of work required before there is a real understanding of
the effects of a full NNLO treatment; not the least of the work 
is in the onerous
task of deriving all the necessary NNLO terms. It may still be
premature~\cite{misc:DIS2000:Ellis} to worry too much about the
somewhat strange behaviour of the NNLO gluon density and $F_L$,
until a full NNLO treatment is possible. Nevertheless, the increased
precision of the data becoming available and the rapid theoretical developments
combine to make the subject of global PDF fitting and structure functions
both topical and interesting.
\subsection{$F_2$ and its derivatives}
\label{sec:sat}
\begin{figure}[h]
\begin{center}
\epsfig{file=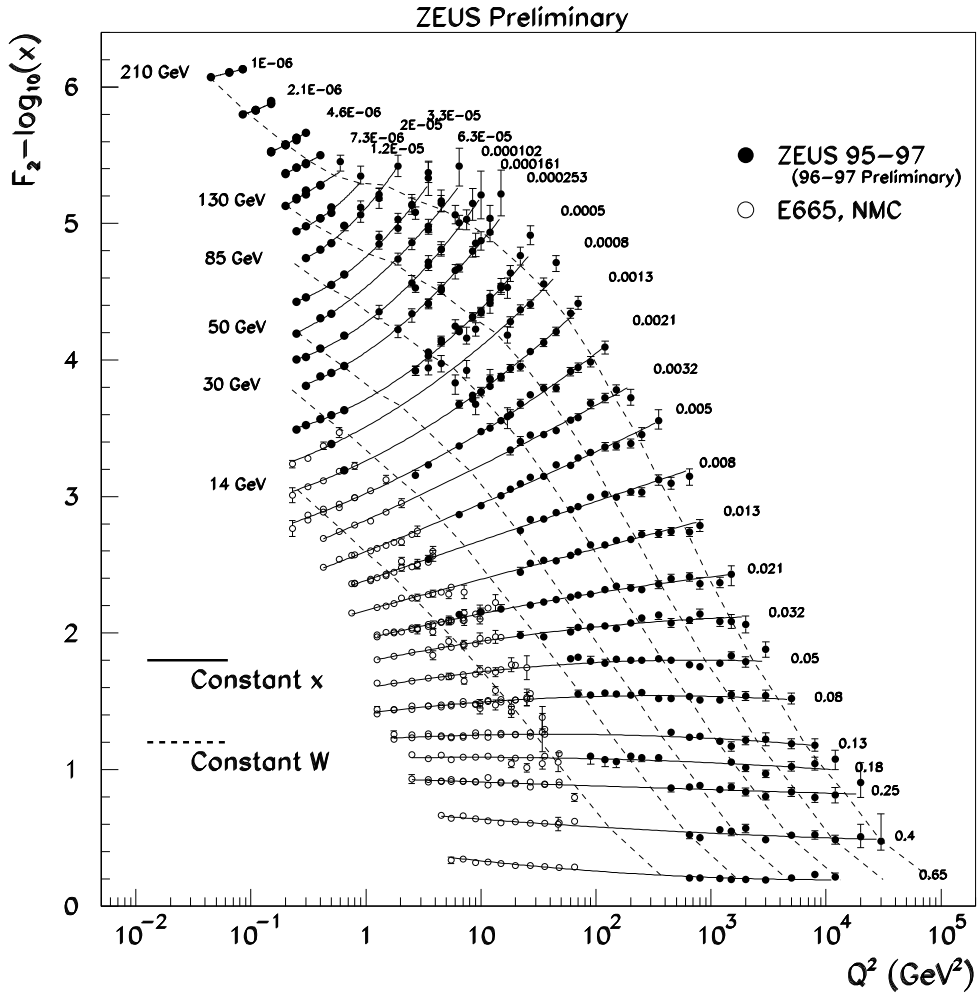,%
      width=15cm,%
      height=15cm%
        }
\end{center}
\caption{Compilation of ZEUS $F_2$ data, both published and preliminary
from the 1996-97 data sample in $x$ bins as a function of $Q^2$. Each
$x$ bin is shifted by an additive constant for ease of visibility.
Data from NMC and E665 is also shown.
The dotted lines show lines of constant $W$, while the solid lines are
fits to the form of equation~\protect\ref{eq:F2param}.} 
\label{fig:ZEUS:6OF2}
\end{figure}

With the publication of the final data from the very low-$(Q^2,x)$ region
measured with the Beam Pipe Tracker (BPT)~\cite{hep-ex-0005018} 
as well as the latest
high-precision $F_2$ data, ZEUS now has precise data over a 
remarkable six orders of magnitude in $x$ and $Q^2$. 
This data is shown in $x$ bins as a function of $\ln Q^2$ 
in figure~\ref{fig:ZEUS:6OF2}, together with fixed target data
from NMC and E665, which extends the range in the direction of medium
$x$ and $Q^2$. 

The availability of this very wide range of precise data makes possible qualitatively new investigations of models that describe $F_2$. As discussed
in section~\ref{sec:f2med}, the logarithmic derivative of $F_2$ is directly
proportional to the gluon density, which in turn is by far the dominant parton
density at the small values of $x$ of interest here. It is therefore interesting 
to examine the behaviour of such logarithmic derivatives as a function of both
$x$ and $Q^2$. Plots of $\partial F_2/ \partial \ln Q^2$
as a function of $x$ were first presented for the ZEUS data by 
Caldwell at the DESY Theory Workshop in 1997 and subsequently published
by ZEUS~\cite{epj:c7:609}, and led to much comment in the literature.
The range and quality of the data available at the time meant that severe
restrictions were placed on how the data could be binned and parameterised.
These restrictions led to several erroneous suggestions that the features of
this plot were a consequence of trivial kinematics. The quality and range
of the currently available data now permits a much better-defined procedure
to be followed in constructing plots of the logarithmic derivative.

The data shown in figure~\ref{fig:ZEUS:6OF2}, particularly in the lower-$x$
bins, are clearly inconsistent with a linear dependence on $\ln Q^2$, as was
pointed out for the preliminary H1 data by Klein~\cite{misc:kleinlps99}. 
The solid curves
on the figure correspond to fits to a polynomial in $\ln Q^2$ of the form:
\begin{equation}
F_2 = A(x) + B(x) \left(\log_{10} Q^2\right) + C(x) 
\left(\log_{10}Q^2\right)^2  
\label{eq:F2param}
\end{equation}
which gives a good fit to the data through the entire kinematic range. The dotted lines on figure~\ref{fig:ZEUS:6OF2} are lines of constant $W$.
The curious `bulging' shape of these contours of constant $W$ in the
small-$x$ region immediately implies that something interesting is
going on there. Indeed, simple inspection of figure~\ref{fig:ZEUS:6OF2} shows
that the slope of $F_2$ at constant $W$ begins flat in the scaling region,
increases markedly as the gluon grows and drives the evolution of $F_2$
and then flattens off again at the lowest $x$. 

\begin{figure}[h]
\begin{center}
\epsfig{file=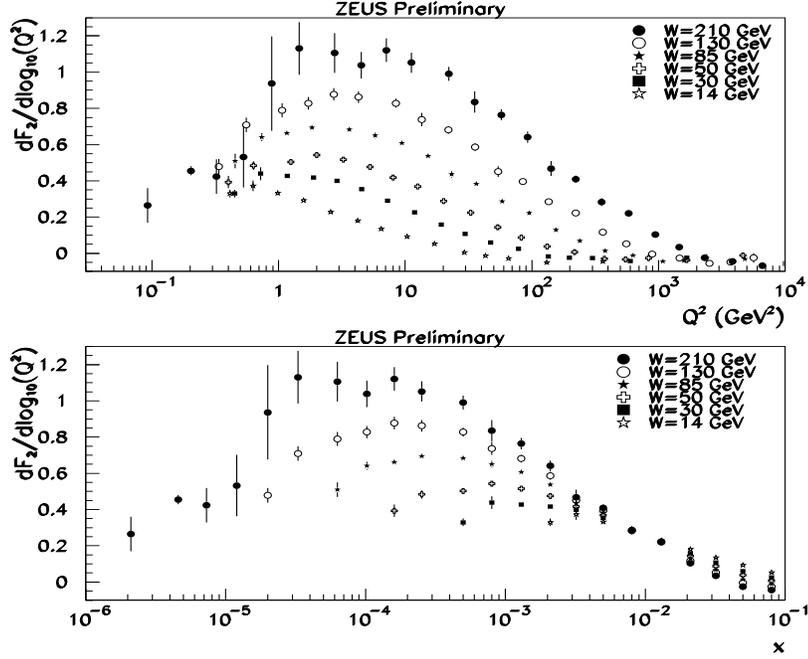,%
      width=12cm,%
      height=10cm%
        }
\end{center}
\caption{The logarithmic derivative of the ZEUS $F_2$ data in six bins
of $W$, plotted as a function of $Q^2$ and $x$.}
\label{fig:ZEUS:logder}
\end{figure}

This behaviour is made clear and explicit in figure~\ref{fig:ZEUS:logder}, which
shows the logarithmic derivative evaluated at $(x, Q^2)$ points
along the contours of fixed $W$
shown on figure~\ref{fig:ZEUS:6OF2} according to the derivative of
equation~\ref{eq:F2param}, {\it viz.}:
\begin{equation}
\frac{\partial F_2}{\partial \log_{10} Q^2} = B(x) + 2C(x) \log_{10} Q^2  
\label{eq:F2deriv}
\end{equation}
where the data are plotted separately as functions of $\ln Q^2$ and $\ln x$.
The data was adjusted where necessary to the appropriate $Q^2,x,W$-bin by 
using the ALLM parameterisation~\cite{pl:b269:465}.
The error bars on the points are evaluated from the errors on the parameters
in equation~\ref{eq:F2param} and consist of the statistical and systematic
uncertainties added in quadrature. The correlations on the errors are, however,
not taken into account, so that the error bars shown are slight over-estimates.

The turn-over in the derivatives in all $W$ bins is marked, and confirms the
similar feature seen in the original ZEUS plot, but now with much better
defined kinematic conditions. When plotted as a function of $\ln Q^2$, the
maximum in the derivative moves to larger $Q^2$ as $W$ increases, while
as a function of $\ln x$, the maximum moves to smaller $x$ as $W$ increases. 

It is of interest to speculate at this point as to what dynamical mechanism
might be causing the behaviour exhibited in figure~\ref{fig:ZEUS:logder}. Since
the logarithmic derivative is proportional at leading order to the gluon density, the obvious inference that can be drawn from the data is that the
rise in the gluon density at low $x$ begins to soften and eventually to fall
as $x$ decreases. Indeed, several indications of such an
effect have been discussed in earlier sections of this talk. Such an effect
is by no means, as will be seen below, necessarily an indication of deviations from GLAP evolution. Nevertheless, such a fall in the gluon density as $x$
falls is a natural consequence of many models of parton saturation or
shadowing, so that it is of interest to explore their features in more
detail at this point. Before beginning however, it is important to
emphasise that the relative emphasis on dipole models in this
talk is not an indication that they are necessarily `correct', or even
necessarily give a better description of the data than other models,
such as the standard twist-two QCD descriptions. Neverthless, they
do have several attractive features, in particular the rather natural
way in which they can lead to a unified description of diffraction and 
deep inelastic scattering, which makes it useful to discuss their
features in some detail here; not least since in general their
concepts are less familiar to the average particle physicist.    
\subsubsection{Dipole models and shadowing}
\label{sec:dipoles}
Dipole models of DIS
have a long history. 
The basic idea is to transform the `normal' way of looking at DIS, which
considers a virtual photon to be emitted from the incoming lepton and to
collide with a parton emanating from the proton, by transforming to a
topologically equivalent process in which the virtual photon splits into
a quark-antiquark pair. These two descriptions are related by a Lorentz 
transform, since the `normal' view of partons evolving inside the proton
is appropriate to a frame such as the Breit
frame or the infinite-momentum frame, whereas the dipole picture is more
appropriate to the rest frame of proton. In the rest frame of the proton,
the virtual photon splits into a quark-antiquark pair, or dipole, well
downstream of the proton. The formation time of the dipole 
in the proton rest frame is related to the uncertainty in the energy of the
pair by $\tau_{q\overline{q}} \sim 1/\Delta E$, which, in the limit
of small $x$ becomes~\cite{forshaw:1997:pomeron} $\tau_{q\overline{q}} \sim 1/ (xM_p)$, where $M_p$ is the
proton rest mass. Since the distance between the formation
of the dipole and the interaction with the proton implied by
this lifetime is much larger than the proton radius, the
transverse size of the dipole can be considered fixed during the
interaction. Thus, for small $x$, the deep inelastic process can be considered semi-classically as the coherent interaction of the dipole with the
stationary colour field of the proton a long time after
the formation of the dipole. In such a frame the dipole
does not evolve a complex parton structure, which is considered to take
place inside the proton.   

It is clear that the formulation of DIS in
this dipole picture provides a direct link between the processes of deep
inelastic scattering and diffraction. The fully inclusive structure functions
sum over all possible exchanges between the dipole and the proton, dominantly
one- and two-gluon exchange, whereas diffraction is produced by the exchange
of 2 gluons in a colour-singlet state. This deep connection between these
two processes leads to non-trivial predictions which do indeed seem to be
borne out by the data. They have been investigated by several
authors, including Golec-Biernat and W\"{u}sthoff~\cite{pr:d60:114023} and Buchm\"{u}ller, Gehrmann and 
Hebecker~\cite{np:b537:477}. 

\begin{figure}[h]
\begin{center}
\epsfig{file=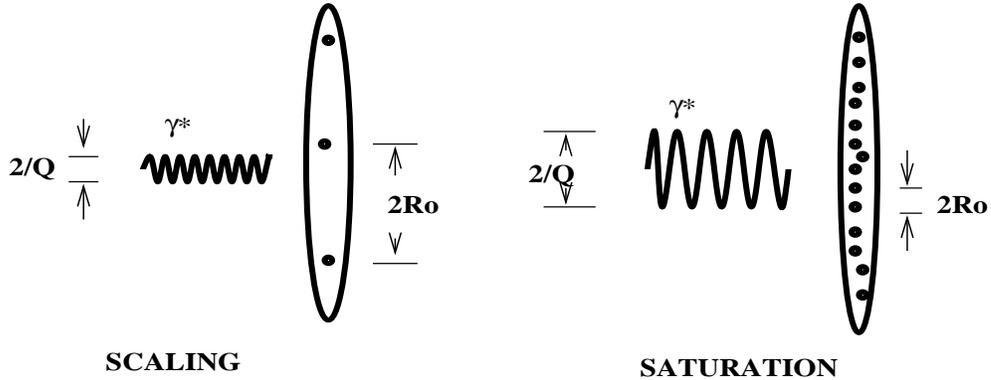,%
      width=13cm,%
      height=5cm%
        }
\end{center}
\caption{Schematic view of the `scaling' and `saturation'
regimes in DIS. The relative sizes of the dipole (proportional to 1/$Q$) and
the mean separation between partons control the behaviour.}
\label{fig:rikdipole}
\end{figure}

Qualitatively, the interaction of the dipole with the colour field of the proton
will clearly depend on the size of the dipole, which is proportional
to $Q^{-1}$. If the separation of the quark
and antiquark is very small, the colour field of the dipole will be effectively
screened and the proton will be essentially `transparent' to the dipole.
At large dipole sizes, the colour field of the dipole is large and it interacts
strongly with the target and is sensitive both to its structure and size.
Such considerations lead naturally to some qualitative understanding of
the process of deep inelastic scattering and saturation illustrated in
a simple one-dimensional model in figure~\ref{fig:rikdipole}.
Here
a one-dimensional distribution of partons inside the proton is considered
in two limiting cases. In the first, labelled as `scaling', the typical
size of the probing dipole is much smaller than the mean separation of the
partons, $R_0$, so that the probability of
interaction is given by the ratio between the
mean size of the dipole and the mean separation of the partons, i.e.\
$1/(QR_0)$. The cross section is thus proportional to
$1/(QR_0)^2$, so that the structure function is independent of
$Q^2$. In the other
case, labelled as `saturation', the size of the dipole is large compared
to the mean separation of the partons, in which case the size which determines
the interaction probability is simply the size of the probe. Thus, for a given
$Q$, the cross section `saturates' to
a constant value.  More generally, when the parton
density is such that the proton becomes `black' and 
the interaction probability is unity, the dipole cross section
saturates for all $Q$ and hence the structure function
becomes proportional to $Q^2$. 
In the Breit-frame-like picture, this is equivalent
to a situation in which the individual partons become so close that they
have a significant probability of interacting with each other before 
interaction with the probe. In the case of gluons, such interactions lead to
two $\rightarrow$ one branchings and hence a reduction in the gluon density.  
Such a picture was the basis for many of the early developments in this
area, in particular the formulation of the modified GLAP evolution
equation including absorptive effects by Gribov, Levin and Riskin~\cite{prep:100:1} and Mueller and Qiu~\cite{np:b268:427},
as embodied in the GLR equation. 
 
In order to look somewhat more quantitatively at the implications
of these ideas, it is necessary to specialise to a particular
model. The dipole
model of Golec-Biernat and W\"{u}sthoff~\cite{pr:d59:014017,pr:d60:114023} 
(G-B\&W) is selected to be discussed in detail. This
does not of course imply that many other 
models~\cite{zfp:c49:607,zfp:c53:331,pl:b326:161,np:b493:354,pl:b425:369,np:b539:535,pr:d60:074012,hep-ph-0007257,np:b415:373,np:b425:471,np:b476:203,np:b487:283,np:b537:477}
both older and more recent, are not equally capable of describing the data. 
In particular, most dipole models share many of the characteristics of the 
G-B\&W model, at least at the rather broad-brush level appropriate to this discussion. 

The interaction between a dipole with a definite transverse
separation 
${\bf \xi}$ at a fixed impact parameter 
${\bf b}$ and the proton can be considered very 
generally~\cite{hep-ph-9911289} in terms of an $S$-matrix element 
$S({\bf \xi},{\bf b})$. The form assumed for this $S$-matrix element 
contains the basic physics of the model in question. In the G-B\&W
model, it is assumed that the impact-parameter dependence can be
factorised and integrated over and that the remaining dependence on
the separation can be approximated by a Gaussian. Explicitly, the cross section
for transverse and longitudinal photon is given by:
\begin{equation}
\sigma_{T,L}(x,Q^2) = \int d^2{\bf r} \int_0^1 dz   
\vert \Psi_{T,L}(z,{\bf r}) \vert^2 \hat{\sigma}(x,r^2)
\label{eq:GBW:sigmatl}
\end{equation}
where $\Psi_{T,L}$ are the light-cone wave-functions for the photon, which
are functions of the fractional momentum of the virtual photon
taken by quark, $z$, and the separation
of the quark and antiquark ${\bf r}$. The photon 
wave-function has the form:
\begin{equation}
\vert \Psi_{T}\,(z,{\bf r}) \vert^2  =
\frac{3 \alpha}{2 \pi^2} \sum_{f} e_f^2 
\left[ \left\{z^2+(1-z)^2\right\} \epsilon^2  K_1^2 (\epsilon r) 
+ m_f^2 K_0^2 (\epsilon r) \right]
\label{eq:GBW:psit}
\end{equation}
\begin{equation}
\vert \Psi_{L}\,(z,{\bf r}) \vert^2  =
\frac{3 \alpha}{2 \pi^2} \sum_{f} e_f^2 
\left[ 4Q^2 z^2 (1-z)^2 K_0^2 (\epsilon r) \right]
\label{eq:GBW:psil}
\end{equation}
where $K_0$ and $K_1$ are McDonald functions and 
\begin{equation}
\epsilon^2 \:=\:  z\,(1-z)\,Q^2\,+\,m_f^2
\label{eq:GBW:epsilon}
\end{equation}
where $m_f$ is the mass of the quark in the dipole. Neglecting for the moment
the fermion mass, the fact that the argument of $K_0$ and $K_1$ is
$\epsilon r$ implies that the `effective' size of the dipole
configuration is proportional to $1/\{ Q \sqrt{z(1-z)}\}$.
Thus, the fact that the longitudinal wave-function from 
equation~\ref{eq:GBW:psil} 
is proportional to $z(1-z)$, whereas the transverse
wave-function in equation~\ref{eq:GBW:psit} is proportional to $z^2+(1-z)^2$ 
implies that the larger configurations, when $z$ or $(1-z) \rightarrow 0$,
are suppressed for the longitudinal photons~\cite{misc:forshaw:priv}. For large dipole configurations, the integral over $z$ in equation~\ref{eq:GBW:sigmatl} picks up contributions only from the end-points, in which either the
quark or the antiquark carries essentially all of the photon momentum;
such configurations are therefore known as `aligned'.
Since the colour
field and hence the interaction probability is lower for smaller dipoles,
the dipole cross section is dominated in most areas of phase space by the transversely polarised component of the virtual photon. 

The sub-process cross section, $\hat{\sigma}$, in equation~\ref{eq:GBW:sigmatl}
is related to the $S$-matrix element discussed above, and is assumed in
the G-B\&W to have the form:
\begin{equation}
\hat{\sigma}\,(x,r^2)\;=\; \sigma_0\; g\,(\hat{r}^2)
\label{eq:GBW:sigmahat}
\end{equation}
where:
\begin{equation}
g\,(\hat{r}^2)\;=\;1\,-\,e^{-\hat{r}^2}
\label{eq:GBW:g}
\end{equation}
\begin{equation}
\hat{r}\;=\; \frac{r}{2\,R_0(x)}
\label{eq:GBW:rhat}
\end{equation}
and:
\begin{equation}
R_0\,(x)\,=\, \frac{1}{Q_0}\;\left( \frac{x}{x_0} \right)^{\lambda/2}
\label{eq:GBW:R0}
\end{equation}
These definitions contain the essential dynamics of the G-B\&W model. At large
`rescaled' dipole sizes, 
$\hat{r}$, $g \rightarrow$ constant and the cross section
saturates. For small $\hat{r}$, the cross-section increases quadratically
with $\hat{r}$, which, from equations~\ref{eq:GBW:rhat} and 
\ref{eq:GBW:R0}, implies
an $x^{-\lambda}$ rise as seen in the data. In order to pick up the $Q^2$
dependence it is necessary to do the integral in equation~\ref{eq:GBW:sigmatl}
for the transverse component. At small $\epsilon r$, the McDonald functions
can be approximated by:
\begin{eqnarray}
K_0(\epsilon r) & \sim & \ln \frac{1}{\epsilon r} 
\label{eq:GBW:K0}\\
K_1(\epsilon r) & \sim & \frac{1}{\epsilon r}
\label{eq:GBW:K1}
\end{eqnarray}
while at large $\epsilon r$ they are exponentially
suppressed. Thus, it is clear that the dominant contribution to the integral comes
from the $K_1$ term for $\epsilon r < 1$. 
This corresponds to the small $\hat{r}$ case discussed above so 
that for `small' dipoles, i.e. $r < 1/Q << R_0$, 
for which $\epsilon r < 1$ is automatically satisfied, the saturation
radius becomes:
\begin{equation*}
\hat{\sigma} \sim \frac{\sigma_0 r^2}{R^2_0}
\end{equation*}
Substituting in equation~\ref{eq:GBW:sigmatl} using equation~\ref{eq:GBW:K1}
the integral collapses to:
\begin{equation}
\sigma_T \sim 
\frac{\sigma_0}{R_0^2}\,  
\int_0^1 (z^2 + (1-z)^2) dz
\int_0^{1/Q^2} d r^2\,\epsilon^2\,
\left(\frac{1}{\epsilon^2 r^2}\right) r^2
\propto  \frac{1}{Q^2}
\frac{\sigma_0}{R^2_0}
\label{eq:GBW:sigmatsmallr}
\end{equation}
since the $z$ integral can be factored out since the $\epsilon$ terms cancel 
and the $r$ integral is limited to an upper limit of $1/Q$ by construction.
At constant $x$, therefore, $F_2$ ($\propto Q^2 \sigma^{\gamma^*p}$) exhibits
scaling. By analogy it is easy to see that for `small' dipoles in which
the characteristic size $1/Q > R_0$, the integral
in equation~\ref{eq:GBW:sigmatsmallr} must be split into two parts, in
which $\hat{\sigma}$ is quadratic in $r$ for small values of $r$
and constant for large $r$, i.e. 
\begin{equation}
\sigma_T \sim   \int\limits_0^{R_0^2} d\,r^2\,
\left(\frac{1}{r^2}\right) \sigma_0 \frac{\,r^2}{R^2_0} + 
\int\limits_{R_0^2}^{1/Q^2} d r^2 
\left(\frac{1}{r^2}\right) \sigma_0
\propto \sigma_0 + 
\sigma_0 \ln\left(\frac{1}{Q^2\,R_0^2}\right)
\label{eq:GBW:sigmatlarger}
\end{equation}
which predicts that $F_2$ is proportional to $Q^2$ (modified by a slow
logarithmic dependence). Thus it can be seen that the presence of
an additional length scale in the problem, the saturation radius, leads
to the prediction that for sufficiently large dipoles (i.e. small $Q^2$),
$F_2$ will become proportional to $Q^2$. The boundary between these
two types of behaviour is the so-called `critical line' (given
by $1/Q = R_0(x)$), which clearly depends on $x$. With
increasing $W$, the transition occurs for smaller $x$ and larger
$Q^2$. 

\begin{figure}[h]
\begin{center}
\epsfig{file=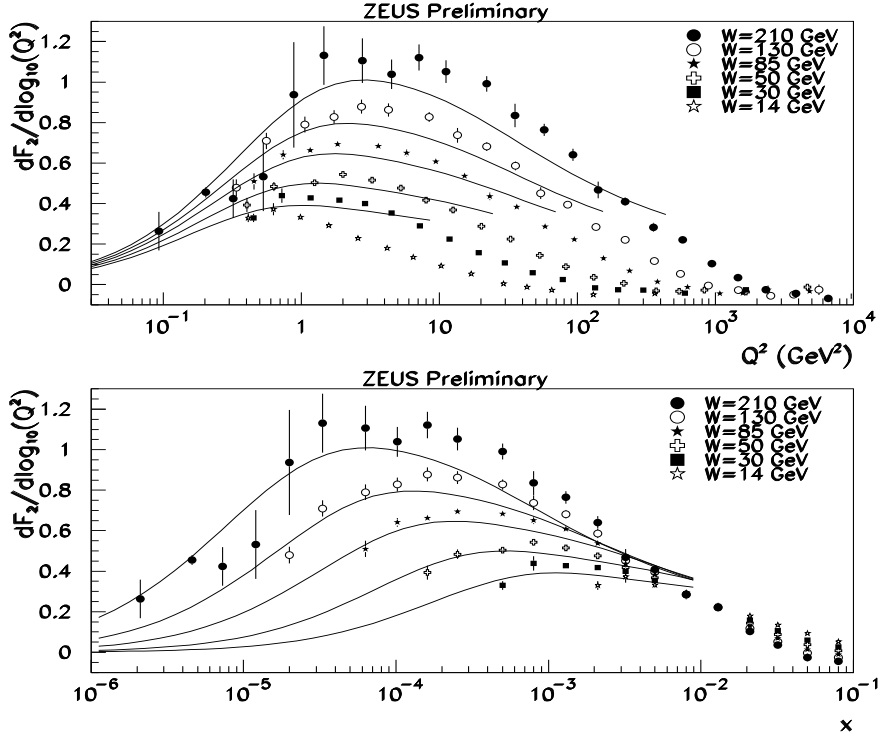,%
      width=12cm,%
      height=10cm%
        }
\end{center}
\caption{Curves showing the G-B\&W model predictions for the logarithmic
derivatives of $F_2$ in bins of constant $W$ as a function of
$Q^2$ and $x$, compared to the ZEUS data. The curves are only plotted
for $x < 10^{-2}$, the limit of validity of the model.}
\label{fig:GBW:logder}
\end{figure}

Having given a broad-brush overview of the main implications of the
G-B\&W model, it's predictions for the logarithmic slope of $F_2$ can be
investigated. The more detailed treatment in Golec-Biernat and
W\"{u}sthoff~\cite{pr:d59:014017} modifies the conclusions of
equations~\ref{eq:GBW:sigmatsmallr} and \ref{eq:GBW:sigmatlarger} by the
inclusion, among other factors, of `large' dipole pairs together
with the longitudinal contribution, to give:
\begin{equation} 
\sigma^{\gamma^* p}(x,Q^2) = \sigma'_0 \left\{
\left(\frac{x_0'}{x}\right)^{\lambda'}\frac{Q_0^2}{Q^2} \ln\left[  
\left(\frac{x}{x_0'}\right)^{\lambda'} \frac{Q^2}{Q_0^2}+1\right]
+ \ln\left[ \left(\frac{x_0'}{x}\right)^{\lambda'} \frac{Q_0^2}{Q^2}+1\right]
\right\}
\label{eq:GBW:sigmatot}
\end{equation}
where the primes denote that the constants are to be optimised by a fit
to the available data. The salient characteristics of 
equations~\ref{eq:GBW:sigmatsmallr} and \ref{eq:GBW:sigmatlarger} remain,
although equation~\ref{eq:GBW:sigmatsmallr} has acquired a logarithmic
modification. The first term therefore governs the behaviour at high
$Q^2$, while the second term is dominant at low $Q^2$. Multiplying
equation~\ref{eq:GBW:sigmatot} by $Q^2$ to convert it to $F_2$ and
taking the logarithmic derivative leads to: 
\begin{equation*}
\frac{\partial F_2}{\partial \ln Q^2} \sim x^{-\lambda'}
\end{equation*}
for high $Q^2 (>> 1/R_0^2$) and to the derivative acquiring a term
proportional to
\begin{equation*}
-Q^2 \sigma'_0
\end{equation*}
for low $Q^2 (\sim 1/R_0^2$) at fixed $x$, thereby reducing the 
size of the derivative. 
Thus the expected power-law growth at low $x$ is seen for high $Q^2$,
where the logarithmic derivative in LO GLAP treatment is proportional
to the gluon density,
while at small $Q^2$ the leading behaviour
of both the derivative and $F_2$ becomes proportional to $Q^2$. This
implies a maximum in the logarithmic derivative, as seen in the data 
of figure~\ref{fig:ZEUS:logder}. Qualitatively, therefore, the G-B\&W model
can describe the ZEUS data, as shown by figure~\ref{fig:GBW:logder},
which contains the curves from the original Golec-Biernat and 
W\"{u}sthoff publication, which indeed
is qualitatively in agreement with the ZEUS data. In particular, the
movement of the maximum with $Q^2$ and $x$ as $W$ changes is
quite well reproduced, but there are clear differences,
particularly at higher $Q^2$ and lower $W$. These are in
regions in which the model has known problems, and it could
well be that a fit to the ZEUS data, which were not available at the
time of the original paper, would improve the agreement.

\begin{figure}[h]
\begin{center}
\epsfig{file=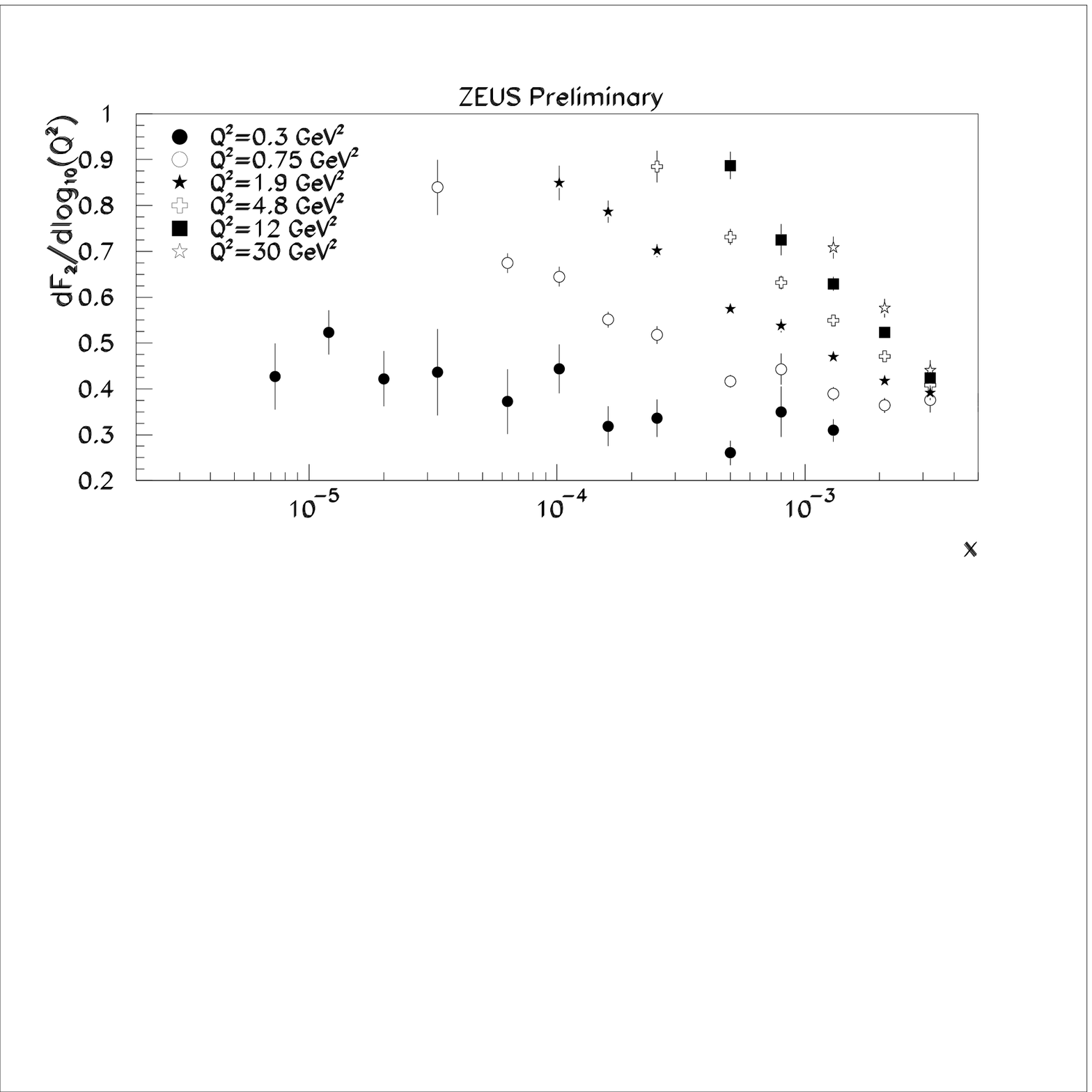,%
      width=12cm,%
      height=7cm,%
      clip=%
        }
\end{center}
\caption{The logarithmic
derivatives of the $F_2$ data in bins of constant $Q^2$ as a function of
$x$.}
\label{fig:ZEUS:logder:fixedQ2}
\end{figure}

It is also of interest to examine the logarithmic derivative at fixed $Q^2$
rather than fixed $W$ as a function of $x$. This is shown for the ZEUS data
in figure~\ref{fig:ZEUS:logder:fixedQ2}. 
The derivative is relatively
straight as a function of $\ln x$, and exhibits a slow change with $Q^2$
at larger $Q^2$, which becomes rapid for $Q^2 < 2$ GeV$^2$. This behaviour
is simply a reflection of the `valence-like' gluon behaviour at low $Q^2$
in which the gluon density in NLO QCD fits falls rapidly to zero while
the sea remains non-zero. It is also reproduced by the G-B\&W model.
The logarithmic derivative of equation~\ref{eq:GBW:sigmatot} multiplied
by $Q^2$ at fixed  $Q^2$ shows that the slope changes from being proportional
to $x^{-2\lambda'}$ to $\ln x^{-\lambda'}$, so that there is simply a change
in slope rather than a turn-over. Moreover, the G-B\&W `critical line',
which predicts the position of the transition to saturation behaviour,
is much steeper in $x$ than in $Q^2$, so that the transition point 
for fixed $Q^2$ is
generally at an $x$ outside the kinematic region of the data. The exception is
the $Q^2 = 0.75$ GeV$^2$ data, where the transition is predicted to
occur at around $x \sim 5 \cdot 10^{-3}$. Such a change in slope
can certainly not be ruled out by the data. 

\begin{figure}[h]
\begin{center}
\epsfig{file=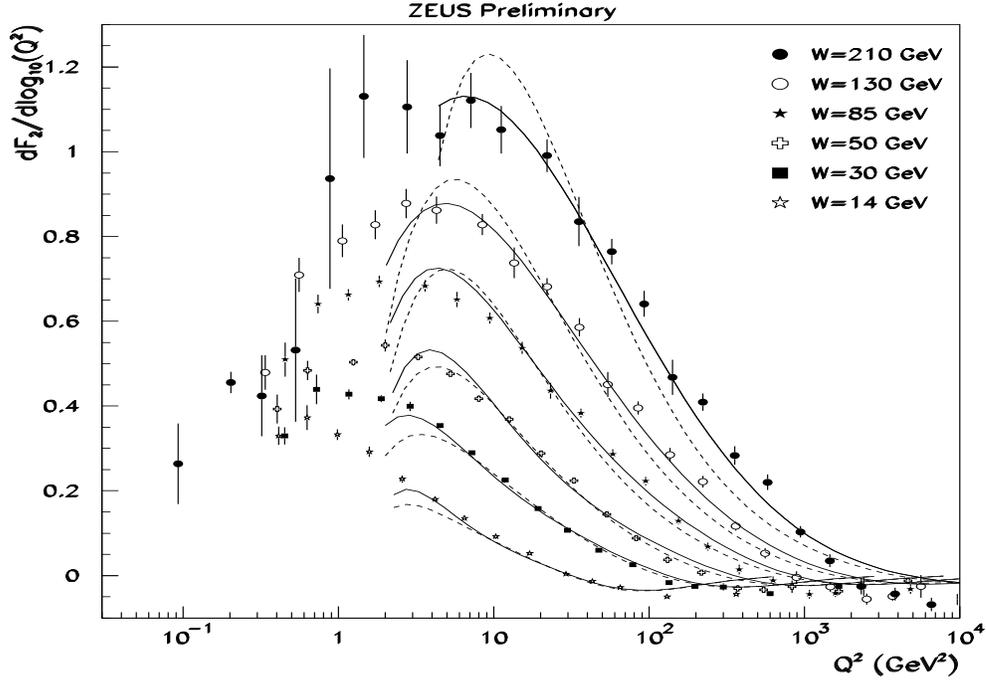,%
      width=13cm,%
      height=9cm%
        }
\end{center}
\caption{The logarithmic
derivatives as a function of
$Q^2$ compared to the predictions of the latest MRST fits
(dotted line)
as well as the modification including the LO BFKL kernel by
Thorne (solid line).}
\label{fig:ZEUS:logder:MRST}
\end{figure}

Although the qualitative agreement of the G-B\&W
saturation model with the ZEUS data is intriguing,
there are many other possible explanations. Several other
saturation and/or dipole models can describe the general
trends of the data~\cite{misc:forshaw:priv,misc:maor:osaka}.
It was already
remarked that the parameterisation of Haidt of equation~\ref{eq:F2:Haidt}
also gives rise to a turnover in the logarithmic derivative. Furthermore,
several NLO QCD analyses seem able to reproduce the turn-over and
other general features of the data. Bl\"{u}mlein~\cite{misc:bluemlein:priv}
has used the GRV framework to fit qualitatively the ZEUS data. 
Roberts~\cite{misc:roberts:priv} has produced logarithmic derivatives
using the MRST fits; although they do produce a turn-over, its
position and the lower $Q^2$ and $x$ slopes do not agree well with the
data. This is scarcely surprising since the whole parton picture must 
already be questionable in such a kinematic region, and there
may well be important higher-twist contributions. Nevertheless, 
Thorne~\cite{misc:thorne:priv}, including the BFKL-motivated
modification of the splitting functions discussed in
*section~\ref{sec:interpretation}\ref{sec:global}, 
section~\ref{sec:global}, 
has produced modified MRST fits that
give an improved fit to the ZEUS data. Both
results are shown in figure~\ref{fig:ZEUS:logder:MRST}.
 
From the above it is clear that the current ZEUS data can certainly
not be used to claim evidence for saturation effects in the HERA
kinematic range. Higher-precision data will certainly help in
distinguishing competing explanations. However, it does not seem
likely that the existence or otherwise of parton saturation can
be unambiguously established at HERA, at least from studies of the
logarithmic derivatives of $F_2$ alone. 
One problem is that the centre-of-mass energy
of HERA means that the interesting areas at low $x$
in which the saturation effects become large is necessarily at 
$Q^2 < 5 - 10$ GeV$^2$, where complications from higher-twist effects 
are inevitable and indeed the whole parton picture at some point ought 
to break down. The only way to improve the situation would be to move 
to a higher energy machine - for instance the proposed THERA option of
colliding TESLA and HERA~\cite{misc:kleinlps99}, or the LEP-LHC $ep$ option. At
THERA the interesting $x$ area for saturation effects would occur 
at $Q^2 > 10$ GeV$^2$. Another possible way forward is to look
simultaneously at several processes, for example DIS and 
inclusive diffraction~\cite{pr:d60:114023} or DIS and elastic $J/\psi$
production~\cite{misc:maor:osaka}.
 
\begin{figure}[h]
\begin{center}
\epsfig{file=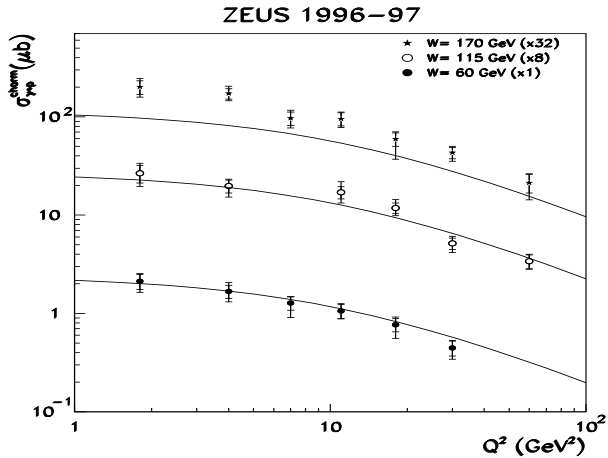,%
      width=8cm,%
      height=6cm%
        }
\end{center}
\caption{The $F_2^{c\overline{c}}$ data expressed
as a total virtual photon-proton cross section
in bins of constant $W$ as a function of
$Q^2$. The lines show the predictions of the G-B\&W model}
\label{fig:GBW:ZEUS:F2charm}
\end{figure}

Interestingly, the G-B\&W model for DIS charm production does
predict a turn-over in the logarithmic derivative at higher values of
$Q^2$. Figure~\ref{fig:GBW:ZEUS:F2charm} shows
a preliminary plot of 
the ZEUS data on DIS charm as a total virtual-photon cross section
at constant $W$ vs $\log Q^2$. 
There is a clear flattening of the derivative
of the cross section in the region somewhat less than 10 GeV$^2$, which is
well reproduced by the G-B\&W prediction. However, this effect is not related
to the saturation of parton densities at low $x$, but rather to
the charm mass and the resultant size of the dipole. In the discussion
above on the G-B\&W model, quark-mass effects were ignored, although they
play an important role, particularly at the lowest values
of $Q^2$, and in ensuring that the cross section matches to the
photoproduction data. The inclusion of fermion mass effects
for instance changes the form of equation~\ref{eq:GBW:sigmatlarger} to
become:
\begin{eqnarray*}
\sigma_T &\sim& \sigma_0 \ln \left( \frac{1}{m_f^2 R_0^2} \right) \\  
\sigma_L &\sim& \sigma_0 \frac{Q^2}{m_f^2}
\end{eqnarray*}
where $m_f$ is the fermion mass and $m_f^2 >> Q^2$. The effect of the charm quark mass is to insert another length scale into the 
problem~\cite{misc:golec:priv}, since the
size of the charm-anticharm dipoles cannot grow beyond the cut-off imposed
by $1/m_c$, as indicated by
equation~\ref{eq:GBW:epsilon}.
It is this length scale which causes the turn-over in the charm data. 
Although not related to saturation effects, the agreement of the 
ZEUS data with the G-B\&W prediction is a beautiful confirmation 
of the general physics behind the dipole model.

\section{Summary and outlook}
\label{sec:summary}

It is difficult to overestimate the effect which the advent of HERA has had on
the study of low-$x$ physics. It has moved from its infancy at least
to and perhaps even beyond the kindergarten. One key to this development is
of course the vast kinematic range opened up by HERA, but the other is the
careful experimentation of the ZEUS and H1
experiments as well as the enormous efforts of many theoreticians. 
The precision of the data is now driving many theoretical investigations. Of course, these studies are important not only from the point 
of view of understanding the subtleties of QCD. They are of first importance 
in understand the data from future colliders, in particular LHC. 
Knowledge of the data at the kinematic
limits of HERA governs the understanding of
the backgrounds for much of the discovery physics of 
LHC. The question of deviations from GLAP evolution, while
fascinating experimentally and theoretically, is also of crucial importance to
predictions of various SM and exotic processes at LHC. In the next
few years, the centre
of attention at HERA will switch somewhat from low-$x$ to high-$x$ physics,
as the HERA upgrade allows ZEUS and H1 to fulfill their potential
as precision probes of the electroweak sector. Nevertheless, further precision
at HERA and the Tevatron is both possible and desirable in the inclusive
processes, and the greatly increased luminosity of the upgrade will permit
investigations of exclusive processes and difficult areas of phase space which
will permit the fascinating glimpses of possible deviations from GLAP evolution
at low $x$ to be investigated further. This is truly an exciting time
in 
low-$x$ physics. 
     
\section*{Acknowledgements}
I am grateful to E.~Gabathuler, I.~Butterworth and J.~Ellis for
organising a most pleasant but also remarkably relaxed and stimulating
Royal Society discussion meeting.  I am grateful to many colleagues
for stimulating conversations. I am
particularly grateful to U.~Katz for his help with
LaTeX/BiBTeX etc., to J.~Bl\"{u}mlein, 
J.~Forshaw, W.G.~Stirling and R.G.~Roberts 
for throwing light into my theoretical darkness, and to the latter
for proving the MRST fit to the logarithmic slopes; to R.S.~Thorne
and K.~Golec-Biernat who not only enlightened me and produced plots 
included in this writeup, but also read the manuscript carefully and
pointed out the errors of my ways. Finally I want to thank R.~Yoshida,
not only for all of the aforementioned assistance, but also for many
happy hours of discussion on structure functions, dipoles, etc; a
pleasant vacation from the rigours of `managing' ZEUS!

\end{document}